\documentclass[twoside,twocolumn]{article}

\usepackage[sc]{mathpazo} 

\usepackage[T1]{fontenc} 
\usepackage{lmodern} 
\usepackage{microtype} 
\usepackage[version=4]{mhchem}
\usepackage[english]{babel} 

\usepackage{graphicx}
\usepackage{color}
\usepackage[top=2cm,bottom=2cm,left=1.5cm,right=1.5cm,columnsep=20pt]{geometry} 
    {\large\scshape Computational Details%
    \par\medskip\normalfont\normalsize}%
    {}%
\newenvironment{acknowledgement}
    {\large\scshape Acknowledgement%
    \par\medskip\normalfont\normalsize}%
    {}%
    {\large\scshape Supporting Information%
    \par\medskip\normalfont\normalsize}%
    {}%
\usepackage{amsmath}
\usepackage{amssymb}
\usepackage{multirow} 
\usepackage[normal, small,labelfont=bf,up,textfont=it,up]{caption} 
\usepackage{booktabs} 

\usepackage{lettrine} 

\usepackage{enumitem} 
\setlist[itemize]{noitemsep} 

\usepackage{abstract} 

\usepackage{titlesec} 
\renewcommand\thesection{\Roman{section}} 
\renewcommand\thesubsection{\roman{subsection}} 
\titleformat{\section}[block]{\large\scshape\centering}{\thesection.}{1em}{} 
\titleformat{\subsection}[block]{\large}{\thesubsection.}{1em}{} 

\usepackage{fancyhdr} 
\usepackage{titling} 

\usepackage{hyperref} 

\pretitle{\begin{center}\Large\bfseries} 
\posttitle{\end{center}} 

\def\bea{\begin{eqnarray}}
\def\eea{\end{eqnarray}}
\def\ben{\begin{equation}}
\def\een{\end{equation}}
\def\benu{\begin{enumerate}}
\def\enu{\end{enumerate}}

\def\bei{\begin{itemize}}
\def\eei{\end{itemize}}
\def\beit{\begin{itemize}}
\def\eit{\end{itemize}}
\def\benu{\begin{enumerate}}
\def\enu{\end{enumerate}}

\def\n{n}

\def\sss{\scriptscriptstyle\rm}

\def\1var{(\bx_1...\bx\N)}

\def\half{\frac{1}{2}}

\def\br{{\bf r}}

\def\bx{{x}}

\def\x{_{\sss X}}
\def\c{_{\sss C}}
\def\s{_{\sss S}}
\def\xc{_{\sss XC}}

\def\N{_{\sss N}}

\def\HF{^{\rm HF}}

\def\GGA{^{\rm GGA}}

\def\PBE{^{\rm PBE}}

\def\TF{^{\rm TF}}

\def\sph_int{ {\int d^3 r}}

\def\intr{\int d^3r\,}
\def\intrp{\int d^3r'\,}

\usepackage{caption}
\usepackage{graphicx,subcaption}
\usepackage{chemformula}

\title{Density functional analysis: The theory of density-corrected DFT} 

\author{%
\textsc{Stefan Vuckovic, John Kozlowski and Kieron Burke} \\ 
\normalsize Departments of Chemistry and of Physics, University of California, Irvine, CA 92697, USA \\  
\and
\textsc{Suhwan Song and Eunji Sim} \\ 
\normalsize Department of Chemistry, Yonsei University, 50 Yonsei-ro Seodaemun-gu, Seoul 03722, Korea \\
}
\date{\today} 

\renewcommand{\maketitlehookd}{
\begin{abstract}
\noindent 
Density-corrected density functional theory (DC-DFT) is enjoying substantial
success in improving semilocal DFT calculations in  a wide variety of chemical
problems.   This paper provides the formal theoretical framework and assumptions
for the analysis of any functional minimization with an approximate functional.
We generalize DC-DFT to allow comparison of any two functionals, not just comparison
with the exact functional.
We introduce a linear interpolation between any two approximations, and use the
results to analyze global hybrid density functionals.
We define the basins of density-space in which this analysis should apply, and give
quantitative criteria for when DC-DFT should apply. We also discuss the effects of strong correlation on density-driven error, utilizing the restricted HF Hubbard dimer as an illustrative example.
\end{abstract}
}

\def\rv{{\bf r}}

\def\beq{\begin{equation}}
\def\eeq{\end{equation}}
\def\non{\nonumber \\}

\def\half{\frac{1}{2}}

\def\sh{_{\sss S H}}
\def\shxc{_{\sss SHXC}}
\def\D{_{\sss D}}
\def\F{_{\sss F}}

\def\tE{\tilde E}
\def\SH{{\rm SH}}

\begin{document}

\maketitle
\chapter{}
\setcounter{secnumdepth}{2}
\renewcommand\thesubsection{\Alph{subsection}}

\sf
\section{Introduction and background}
\label{intro}

Kohn-Sham density functional theory (KS DFT)~\cite{KohSha-PR-65} is widely popular as an electronic
structure method~\cite{PriGroBurk-ARPC-15}. Despite the proliferation of choices of approximate
functionals,  most calculations use one of a few standard approximations
that have been available for the past twenty years, namely generalized gradient
approximations (GGAs) or global hybrids, with some enhancements, 
such as van der Waals corrections~\cite{GriAntEhrKri-JCP-10}
or range separation~\cite{Sav-INC-96}. While moderately accurate for many useful properties, these functionals suffer from well-known deficiencies, including unbound anions, poorly positioned eigenvalues, 
incorrect molecular dissociation curves, underestimation of reaction barriers, and many
others~\cite{CohMorYan-SCI-08,CohMorYan-CR-12}. Thus the never-ending search for improved functionals.

Over the years, many pioneers have shown in specific cases that use of approximate 
functionals on Hartree-Fock (HF) densities can yield surprisingly accurate results.
This includes the early work of Gordon and Kim for weak forces~\cite{GorKim-JCP-72}, Janesko and Scuseria 
for reaction barriers and other properties~\cite{S92,JanScu-JCP-08}, and the original works of Gill {\it et al.}
testing GGA's and hybrids for main group chemistry that led to the adoption of DFT
for widespread use in chemistry~\cite{GillJohPopFri-CPL-92}.  
Even the prototype of KS-DFT, the X$-\alpha$-method of Slater~\cite{Sla-PR-51}, was designed to yield
approximations to HF potentials, which led to an inconsistency between the associated
energy functional and its derivative, the potential (see Ref.~\cite{GriMenBae-JCP-16} for a recent discussion on this topic).  Analysis of this difficulty
was part of the impetus for the KS paper.

The errors made in DFT calculations were formally separated
into two contributions, a functional error and a density-driven error, thereby yielding
a formal framework in which the two errors could be analyzed independently~\cite{KimSimBur-PRL-13}.  This led to the theory of density-corrected DFT (DC-DFT), which explains the success of the early work, and has provided
a simple procedure for significantly improving the results of
semi-local DFT calculations in many situations.  For example,
for Halogen and Chalcogen weak bonds, which have been used in
databases to train van der Waals functionals, the errors are dominated by density-driven
errors in the semilocal functional, so such databases cannot be used for that purpose without
a correction~\cite{KimSonSimBur-JPCL-19}. In addition to the standard semilocal functionals, it has been  recently shown that in specific situations, the energetic accuracy of other density functionals, such as the nonlocal functionals based on adiabatic connection models~\cite{VucIroWagTeaGor-PCCP-17}, can be greatly improved by using the HF density and orbitals~\cite{FabGorSeiSal-JCTC-16,VucGorDelFab-JPCL-18}.  

Thus, DC-DFT, especially in the form of HF-DFT, in which the Hartree-Fock density is used in 
place of the exact density, is an extremely practical procedure for improving energetics
of abnormal DFT calculations, i.e., those dominated by density-driven errors, but
in which the approximate functional is still highly accurate.   

Here, we give a detailed formal analysis of the differences that arise between the 
self-consistent solutions of two distinct density functionals.  We consider any two
functionals, including the possibility of two different approximations.
Thus, DC-DFT is a special case of this more general analysis.   We also consider other special cases, including the one-electron case, for which we can calculate all the quantities arising from our analysis that require access to the exact functional and the exact density. 
The accuracy of PBE for
the H-atom is due to a spurious cancellation of both
 density and functional errors,
as well as exchange and correlation errors.
We extend our analysis to energy differences, that are of the key importance in chemistry. We also construct measures for the abnormality character of DFT calculations.

\section{Density functional analysis}
\label{sec_intro}

In KS DFT~\cite{KohSha-PR-65,PriGroBurk-ARPC-15,CohMorYan-CR-12,Bur-JCP-12,Bec-JCP-14}, the ground state energy and density of a system with an external potential $v$ are given by:
\begin{equation} \label{eq:gs}
E_v= \min_\n \left\lbrace F[n]+ n \cdot v  \right \rbrace ,
\end{equation}
where $n \cdot v =  \intr n(\br)v(\br)$, and where $F[n]$ is the universal part
of the functional commonly partitioned as:
\beq\label{eq:f[n]}
F[n]=T_{\s}[n]+U_{\sss H}[n]+E_{\xc}[n],
\eeq
$T_{\s}[n]$ is the KS noninteracting kinetic energy functional, $U_{\sss H}[n]$ the Hartree energy and  $E_{\xc}[n]$ is the exchange-correlation (XC) functional, which for practical calculations must be approximated. 
Starting from a given approximate or exact XC functional $E_{\xc}[n]$, we can write the corresponding approximate universal functional as:
\beq \label{eq:tildef}
F[n]=F_{\sh}[n]+E_{\xc}[n],
\eeq
where $F_{\sh}[n]$ is the universal functional within the Hartree approximation, which neglects exchange and correlation effects: $F_{\sh}[n]=T_{\s}[n]+U_{\sss H}[n]$. The total energy functional is then given by:
\beq\label{eq:ei}
E_v[n]= F[n] + n \cdot v,
\eeq
As usual, the corresponding ground state energy is obtained  from the following minimization over all $N$-representable densities:
\beq \label{eq:tildeE}
E_v=\min_n E_v[n],
\eeq
and the density that achieves the minimum in Eq~\ref{eq:tildeE} we denote by $n_v$. We define an energetic measure of any arbitrary density difference from $n_v$ as:
\beq  \label{eq:cdelta}
D_v[\Delta n]=E_v[n_v + \Delta n] - E_v \geq 0,
\eeq 
where Eq.~\ref{eq:tildeE} ensures that $D_v[\Delta n] \geq 0$ for any isoelectronic change in $n_v$ (i.e. $\intr \Delta n(\br) =0$). 
We refer to this as the energetic distance from the minimum.
We can use this measure to say that $n$ is sufficiently close to $n_v$, if:
\beq \label{eq:deltac}
D_v[n-n_v] \leq \Delta_c,
\eeq
provided that  $\Delta_c$ is sufficiently small.

Throughout this work, we encounter simple quadratic density functionals, which correspond to normal forms in algebra, and we write:
\beq
A[\Delta n]= \intrp \intr A(\br,\br') \Delta n(\br) \Delta n(\br'). 
\eeq
To gain more insight into the $D_v[\Delta n]$ functional, we can expand $E_v[n]$  around its minimum  in a Taylor series:~\cite{Ern-PRA-94}
\begin{align}
\label{eq:taylor}
E_v[n_v +\Delta n]= E_v+\half  K_v[\Delta n] + \mathcal{O}(\Delta n^3),
\end{align}
where $\Delta n(\br)= n(\br)-n_v(\br)$, and $K_v(\br,\br')$ is given by:
\beq \label{eq:k}
K_v(\br,\br')=\frac{\delta^2 E_v[n]}{\delta  n(\br') \delta  n(\br)} \bigg|_{n=n_v}.
\eeq
Combining Eqs.~\ref{eq:tildef},~\ref{eq:ei} and~\ref{eq:k}, we can write $K_v(\br,\br')$ as:
\beq \label{eq:Kv}
K_v(\br,\br')=f_{\sh}[n_v](\br,\br')+f_{\xc}[n_v](\br,\br'),
\eeq
where 
\beq \label{eq:s}
f_{\sh}[n](\br,\br')= \underbrace{
\frac{\delta^2 T_{\s}[n]}{\delta  n(\br') \delta  n(\br)}  }_{f_{\sss S}(\br,\br')},
+\frac{1}{|\br-\br'|}
\eeq 
and 
\beq
f_{\xc}[n](\br,\br')=\frac{\delta^2 E_{\xc}[n]}{\delta  n(\br') \delta  n(\br)}
\eeq 
is the static XC kernel. 
Combining now Eqs.~\ref{eq:cdelta} and~\ref{eq:taylor}, for arbitrary and sufficiently small density difference (i.e. $D_v[\Delta n] \leq \Delta_c$), we can write $D_v[\Delta n]$ as:
\beq \label{eq:ctaylor}
D_v[\Delta n]\approx \half K_v[ \Delta n] .
\eeq
For any $\Delta n (\br)$, satisfying $\intr \Delta n (\br)=0$, we define:
\beq \label{eq:nalpha}
\Delta n_\beta (\br)= \beta \Delta n (\br).
\eeq
In this way, we can see {\em how far one can go away from $n_v(\br)$}, in the $\Delta n(\br)$ 'direction', and yet {\em stay within} $\Delta_c$ energetically. Plugging Eq.~\ref{eq:nalpha} into Eq.~\ref{eq:ctaylor}, we can easily find the  $\beta = \beta_c$ parameter at the boundary (i.e., the one satisfying: $D_v[\Delta n_\beta] = \Delta_c$), and it is given by:
\beq \label{eq:beta}
\beta_c=\sqrt{\frac{2 \Delta_c}{K_v[ \Delta n] }}.
\eeq
The principal goal of the theory outlined here is to carefully analyze the origin of the energy difference that arises 
between a pair of different density functionals, when applied to the same system/process. For a given pair of approximate (or one approximate and the other exact) XC  functionals: $E_{\xc}^{(0)}$ and $E_{\xc}^{(1)}$, we define their difference as:
\beq \label{eq:dxc}
\Delta E_{\xc}[n]=E_{\xc}^{(1)}[n]-E_{\xc}^{(0)}[n].
\eeq
For a given $v(\br)$, the difference in the two ground state energies arising from a pair of different functionals is:
\beq \label{eq:ev10}
\Delta E_v =  E_v^{(1)} [n_v^{(1)}]- E_v^{(0)} [n_v^{(0)}].
\eeq
By simply adding and subtracting $E_v^{(1)} [n_v^{(0)}]$ from the r.h.s. of Eq.~\ref{eq:ev10}, we find:
\begin{align} \label{eq:dif1}
\Delta E_v =-D_v^{(1)}[-\Delta n_v]+\Delta E_{\xc}[n_v^{(0)}]
\end{align}
where $\Delta n_v(\br)=n_v^{(1)}(\br) - n_v^{(0)}(\br)$.
Reversing the choice of $1$ and $0$, we also find:
\beq  \label{eq:dif2}
\Delta E_v=\Delta E_{\xc}[n_v^{(1)}]+D_v^{(0)}[\Delta n_v].
\eeq
Given that $D_v^{(j)}\ \geq 0$, Eqs.~\ref{eq:dif1} and~\ref{eq:dif2} dictate the following chain of inequalities:
\beq
\Delta E_{\xc}[n_v^{(1)}] \leq \Delta E_v \leq \Delta E_{\xc}[n_v^{(0)}].
\eeq
\begin{figure}[htb]
\includegraphics[width=0.98\columnwidth]{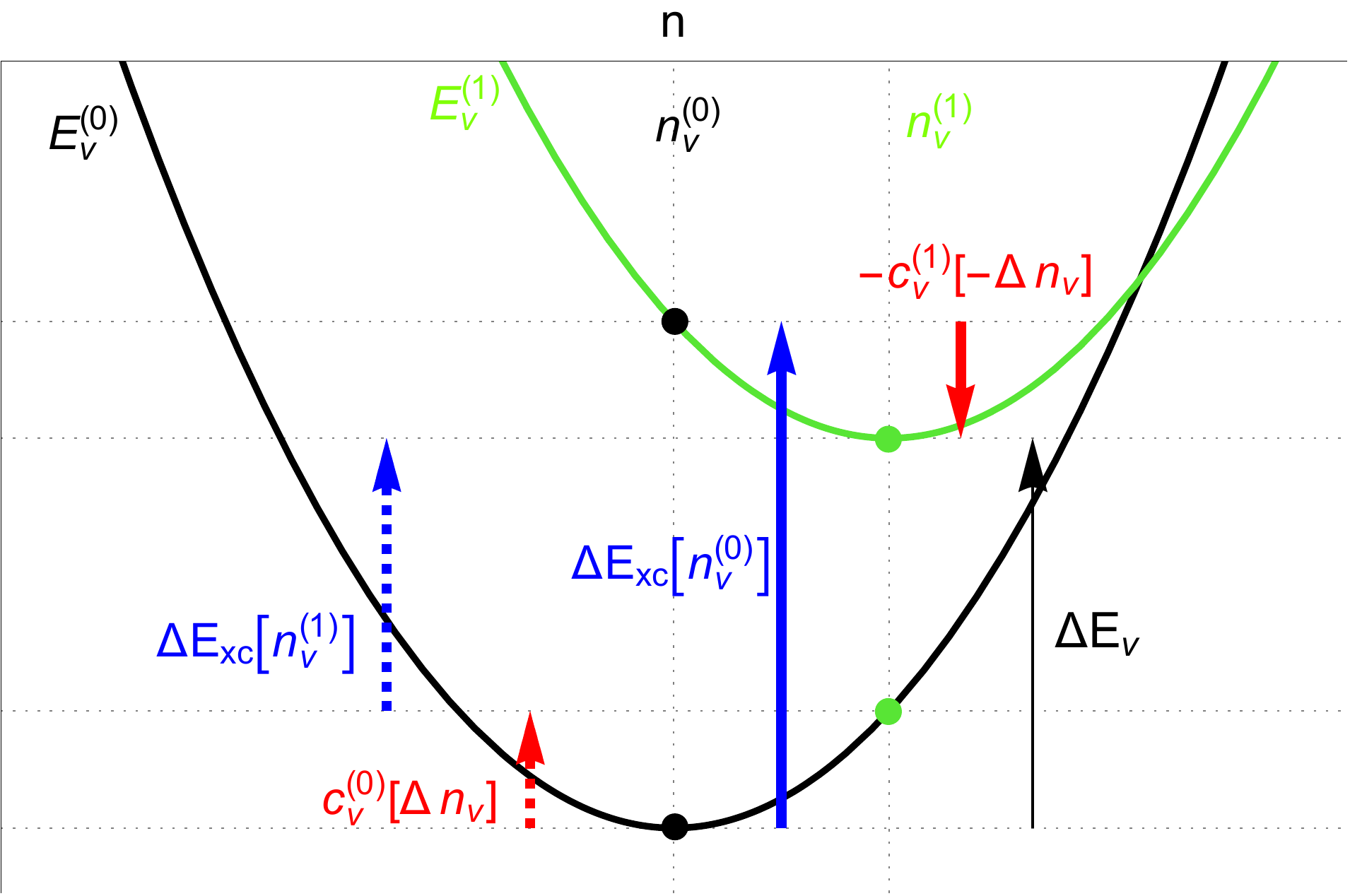}
\includegraphics[width=0.98\columnwidth]{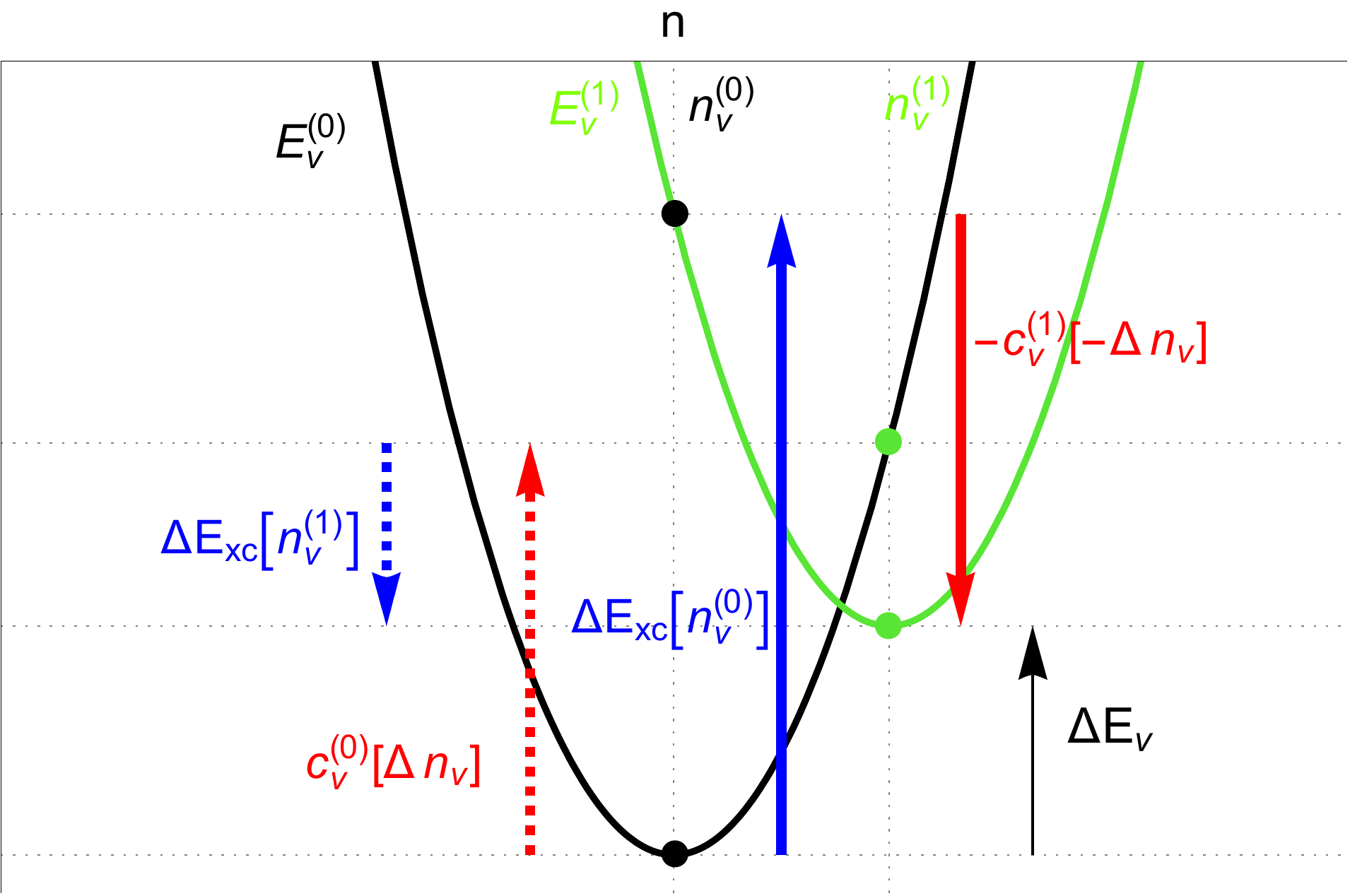}
\caption{Cartoon showing the density-driven and functional-driven contributions
to $\Delta E_v$ (Eqs.~\ref{eq:dif1} and~\ref{eq:dif2}) in an
energy-driven difference (top panel) and a density-driven difference (bottom panel)}
\label{fig:nor}
\end{figure}

By virtue of Eq.~\ref{eq:dxc}, the $\Delta E_{\xc}[n_v^{(i)}]$ quantity
represents the difference between the two functionals evaluated on 
each density. Therefore, we can identify 
$\Delta E_{\xc}[n_v^{(1)}]$ and 
$\Delta E_{\xc}[n_v^{(0)}]$ of Eqs.~\ref{eq:dif1} and~\ref{eq:dif2} as
functional-driven terms. On the other hand, $D_v^{(0)}[n_v^{(1)}]$
and $D_v^{(1)}[n_v^{(0)}]$ are the density-driven terms, as
they are given by the difference between the same energy functional evaluated
on different densities. Generalizing the ideas of
DC-DFT~\cite{KimSimBur-PRL-13,KimSonSimBur-JPCL-19,KimSimBur-JCP-14,
KimParSonSimBur-JPCL-15,WasNafJiaKimSimBur-ARPC-17,
SonKimSimBenHeiBur-JCTC-18,SimSuhBur-JPCL-18}, 
for any pair of density functionals, we classify a
$\Delta E_v $ energy difference as {\em energy-} or {\em density-}driven.
We consider energy-driven $\Delta E_v $ as ones whose functional-driven terms
in Eqs.~\ref{eq:dif1} and~\ref{eq:dif2} strongly dominate the 
density-driven terms (i.e.$\left|\Delta E_{\xc}[n_v^{(0)}]\right|>>
D_v^{(1)}[-\Delta n_v]$ and $\left|\Delta E_{\xc}[n_v^{(1)}]\right|>>
D_v^{(0)}[\Delta n_v]$). On the other hand, in {\it density-driven} cases
the density-driven terms are no longer negligible. In Figure~\ref{fig:nor},
we show the two density-driven and the two functional-driven contributions to their energy difference
in a cartoon representing an energy-driven difference (top panel)
and a density-driven difference (bottom panel). Measures that quantify
density-driven character in a given system (again for a given pair of functionals) will be introduced and discussed in Section~\ref{sec_dif}. 

\section{Density functional interpolation}

To derive an exact expression for $\Delta E_v$ by smoothly connecting $E_{\xc}^{(0)}[n]$ to $E_{\xc}^{(1)}[n]$, we introduce the $\alpha$-parameter dependent XC functional:
\beq \label{eq:exc_alpha}
E_{\xc}^{(\alpha)}[n]=E_{\xc}^{(0)}[n]+ \alpha \Delta E_{\xc}[n].
\eeq
The corresponding total energy functional reads as:
\begin{align} \label{eq:Ealpha}
E_v^{(\alpha)}[n]=  \underbrace{ F_{\sh}[n] + n \cdot v + E_{\xc}^{(0)}[n]}_{E_v^{(0)}[n]} + \alpha \Delta E_{\xc}[n] 
\end{align}
and achieves its minimum at $n_v^{(\alpha)}(\br)$. Thus its ground state energy
is given by $E_v^{(\alpha)}=E_v^{(\alpha)}[n_v^{(\alpha)}]$. More generally, we consider the following energy difference:
\beq \label{eq:ev1alpha}
\Delta E_v^{(\alpha)} = E_v^{(\alpha)} - E_v^{(0)}=  E_v^{(\alpha)} [n_v^{(\alpha)}]- E_v^{0} [n_v^{(0)}].
\eeq
Writing:
\beq \label{eq:Eana}
\Delta E_v^{(\alpha)} = \int_0^\alpha d\alpha'\, \frac{\partial E_v^{(\alpha')}}{\partial \alpha'},
\eeq
from Eq.~\ref{eq:Ealpha}, via the Hellmann-Feynman theorem, it follows:
\beq \label{eq:ealpha}
\frac{\partial E_v^{(\alpha)}}{\partial  \alpha}=\Delta E_{\xc}[n_v^{(\alpha)}]. 
\eeq
Plugging Eq.~\ref{eq:ealpha} into Eq.~\ref{eq:Eana}, we find:
\beq \label{eq:deltaEa}
\Delta E_v^{(\alpha)}  =\int_0^\alpha d\alpha'\, \Delta E_{\xc}[n_v^{(\alpha')}].
\eeq
Equation~\ref{eq:deltaEa} is analogous to, but different from, the adiabatic connection formula for the correlation energy in DFT~\cite{HarJon-JPF-74,LanPer-SSC-75,GunLun-PRB-76}. 
When $\alpha=1$, Eq.~\ref{eq:deltaEa} becomes:
\beq \label{eq:deltaEa1}
\Delta E_v  =\int_0^1 d\alpha\, \Delta E_{\xc}[n_v^{(\alpha)}].
\eeq
This shows that the energy difference between two KS calculations with different XC functionals
can be found knowing only the difference functional and the interpolating ground-state density.
Obtaining $\Delta E_v$ from Eq.~\ref{eq:deltaEa1} requires knowledge of $n_v^{(\alpha)}(\br)$ for all $\alpha$ values between $0$ and $1$. To find $n_v^{(\alpha)}(\br)$, we write the corresponding Euler equation:
\beq \label{eq:euler}
v_{\shxc}^{(0)}[n](\br)
+\alpha \Delta v_{\xc}[n] (\br) + v(\rv)
= \mu.
\eeq
where $v_{\shxc}^{(0)}[n](\br)=\delta F^{(0)}[n] / \delta n(\br)$, and where $\mu$ is the chemical potential. The role of $\mu$ is not relevant here, as we always keep the number of electrons fixed. The density that satisfies Eq.~\ref{eq:euler} is $n_v^{(\alpha)}(\br)$, and by expanding it around $n_v^{(0)}(\br)$: $n_v^{(\alpha)}(\br)= n_v^{(0)}(\br) + \alpha \Delta n_v^{(\alpha)}(\br)$, we can write Eq.~\ref{eq:euler} as:
\begin{align} \label{eq:eulerExpand}
& v_{\shxc}^{(0)}[n_v^{(0)}](\br) +  \alpha  \left[  K_{v}^{(0)}  \cdot  \Delta n_v^{(\alpha)} \right] (\br)  \non
&+ \alpha \left( 
\Delta v_{\xc}[n_v^{(0)}](\br)+ \alpha  \left[   \Delta f_{\xc}\left[n_v^{(0)}\right] \cdot  \Delta n_v^{(\alpha)}   \right]  (\br) 
\right) + v(\br) \non
&= \mu,
\end{align}
where we simplified the notation for the following integral:
\beq
\left[  A \cdot  \Delta n \right] (\br)=\intrp   A (\br,\br') \Delta n (\br').
\eeq
At $\alpha = 0$, Eq.~\ref{eq:euler} becomes:
\beq \label{eq:euler0}
v_{\shxc}^{(0)}[n_v^{(0)}](\br) +v(\br)= \mu
\eeq
Plugging Eq.~\ref{eq:euler0} into Eq.~\ref{eq:eulerExpand} gives:
\begin{align} \label{eq:euler-0}
& 
\left[ K_{v}^{(0)}[n_v^{(0)}]  \cdot  \Delta n_v^{(\alpha)} \right] (\br)    + \Delta v_{\xc}[n_v^{(0)}](\br)  \non 
 & +\alpha \left[ \Delta f_{\xc}[n_v^{(0)}]  \cdot  \Delta n_v^{(\alpha)} \right] (\br)  = 0. 
\end{align}
Also plugging $K_{v}^{(\alpha)}(\br,\br')= K_{v}^{(0)}(\br,\br') + \alpha \Delta f_{\xc}(\br,\br')$ (Eqs.~\ref{eq:Kv} and~\ref{eq:Ealpha}) into Eq~\ref{eq:euler-0}, we obtain:
\begin{align} \label{eq:eulerdelta}
\intrp K_{v}^{(\alpha)}  [n_v^{(0)}] (\br,\br') \, \Delta n_v^{(\alpha)}(\br')= -  \Delta v_{\xc}[n_v^{(0)}](\br).
\end{align}
In principle, $ \Delta n_v^{\alpha}(\br')$ can be obtained by solving Eq.~\ref{eq:eulerdelta}, and we can write the solution in terms of the inverse of $K_{v}^{(\alpha)}[n_v^{(0)}]$:
\begin{align} \label{eq:kinv}
\Delta n_v^{(\alpha)}(\br) =-\intrp  \left\{K_{v}^{(\alpha)}\right\}^{-1}\,
[n_v^{(0)}] (\br,\br')\, \Delta v_{\xc}[n_v^{(0)}](\br').
\end{align}
We expect that $\Delta n_v^{(\alpha)}(\br)$ of Eq.~\ref{eq:kinv} can be fairly approximated by $\Delta n_v$ and this is equivalent to approximating $n_v^{(\alpha)}$ via the following linear interpolation:
\beq \label{eq:lin_interpol}
n_v^{(\alpha)} (\br) \approx n_v^{(0)} (\br)+ \alpha \Delta n_v (\br).
\eeq
To explore in what situation the approximation of Eq.~\ref{eq:lin_interpol} becomes exact, we now write 
$n_v^{(\alpha)}$ as : $n_v^{(\alpha)} (\br)= n_v^{(1)} (\br) - \bar \alpha  \Delta n_v^{(\alpha)'} $, where $\bar \alpha=1 - \alpha$. Repeating the steps given by Eqs.~\ref{eq:euler} to~\ref{eq:kinv}, we find:
\begin{align} \label{eq:kinv1}
\Delta n_v^{(\alpha)'}(\br)=-\intrp  \left\{ K_{v}^{(\alpha)}\right\}^{-1}\,
[n_v^{(1)}] (\br,\br') \, \Delta v_{\xc}[n_v^{(1)}](\br').
\end{align}
In this way, when $\Delta n_v^{(\alpha)'}(\br)$ is equal to $\Delta n (\br)$ of Eq.~\ref{eq:kinv}, then the exact $n_v^{(\alpha)}$ is indeed given by the r.h.s of Eq.~\ref{eq:lin_interpol}. 

To obtain the leading order of $E_v^{(\alpha)}$ in powers of $\alpha$, we set again: $n_v^{\alpha}(\br)=\n_v^{(0)}(\br)+\alpha \Delta n_v^{(\alpha)}(\br)$. Then, $E_v^{(\alpha)}$ becomes: 
\begin{align} \label{eq:Ealphan0deltan}
E_v^{(\alpha)}[n_v^{(\alpha)}]=& F_{\sh}[n_v^{(0)}+\alpha \Delta n_v^{(\alpha)}] + n_v^{(0)} \cdot v  \non
&+ \alpha \Delta n_v^{(\alpha)} \cdot v + E_{\xc}^{(0)}[n_v^{(0)}+\alpha \Delta n_v^{(\alpha)}]  \non
&+ \alpha \Delta E_{\xc}[n_v^{(0)}+\alpha \Delta n_v^{(\alpha)}].
\end{align}
We can expand $E_v^{(\alpha)}[n_v^{(0)}+\alpha \Delta n_v^{(\alpha)}]$ around $n_v^{(0)}(\br)$, and write $E_v^{(\alpha)}$ in powers of $\alpha$: 
\begin{align} \label{eq:Ealpha2}
E_v^{(\alpha)}=& E_v^{(0)}[n_v^{(0)}] +  \alpha \Delta E_{\xc}[n_v^{(0)}] \non
&+\alpha \left( v_{\shxc}^{(0)}[n_v^{(0)}]+ v \right) \cdot \Delta n_v^{(\alpha)} \non
&+ \alpha^2  \left( \Delta v_{\xc}[n_v^{(0)}]  \cdot \Delta n_v^{(\alpha)} + \half  K_v^{(0)}\left[\Delta n_v^{(\alpha)}\right]\right) \non 
&+\mathcal{O}(\alpha^3)
\end{align}
Combining Eqs.~\ref{eq:euler0} and~\ref{eq:Ealpha2}, the third term on the r.h.s of Eq.~\ref{eq:Ealpha2} vanishes:
\begin{align} \label{eq:Ealpha3}
E_v^{(\alpha)}=& E_v^{(0)}[n_v^{(0)}] +  \alpha \Delta E_{\xc}[n_v^{(0)}] \non
&+ \alpha^2
\left( \half K_v^{(0)}\left[\Delta n_v^{(\alpha)}\right]+  \Delta v_{\xc}[n_v^{(0)}] 
  \cdot \Delta n_v^{(\alpha)} \right) \non
&+\mathcal{O}(\alpha^3).
\end{align}
Using Eq.~\ref{eq:eulerdelta}, we can further simplify Eq.~\ref{eq:Ealpha3}:
\begin{align} \label{eq:Ealpha4}
E_v^{(\alpha)} =&  E_v^{(0)}[n_v^{(0)}] +  \alpha \Delta E_{\xc}[n_v^{(0)}] +\frac{\alpha^2}{2} \Delta v_{\xc}[n_v^{(0)}] \cdot \Delta n_v^{(\alpha)} \non 
&+\mathcal{O}(\alpha^3).
\end{align}
From Eq.~\ref{eq:Ealpha4}, we can see that to leading order in $\alpha$, the $\alpha$-dependence of the minimizing energies is quadratic, and it can be found from the xc energies and xc potentials at the endpoints.
We can also use the following functional expansion: 
\begin{align}
\label{eq:delExct}
& \Delta E_{\xc}\left[n_v^{(0)}+\frac{\alpha}{2} \Delta n_v^{(\alpha)}\right] = \non
& \Delta E_{\xc}\left[n_v^{(0)}\right] + \frac{\alpha}{2} \Delta v_{\xc}\left[n_v^{(0)}\right] \cdot \Delta n_v^{(\alpha)} + \mathcal{O}(\alpha^2),
\end{align}
and plug it into Eq.~\ref{eq:Ealpha4} to finally obtain:
\begin{align} \label{eq:fin0}
E_v^{(\alpha)}= E_v^{(0)}\left[n_v^{(0)}\right]  + \alpha \Delta E_{\xc}\left[n_v^{(0)}+\frac {\alpha}{2} \Delta n_v^{(\alpha)}\right] + \ldots
\end{align}
Similarly, writing $n_v^{(\alpha)}=n_v^{(1)}(\br)- \bar \alpha \Delta n_v^{(\alpha)} (\br)$, we can expand $E_v^{(\alpha)}$ around $n_v^{(1)}$ to obtain:
\begin{align}
E_v^{(\alpha)}= E_v^{(1)}\left[n_v^{(1)}\right]  - \bar \alpha \Delta E_{\xc}\left[n_v^{(1)}-\frac {\bar \alpha}{2} \Delta n_v^{(\alpha)}\right] + \ldots
\end{align}
Both results are consistent with applying Eq.~\ref{eq:deltaEa} and expanding only the density to first order in $\alpha$.
In Section~\ref{sec_spec}\ref{sec_one}, we will illustrate the usefulness of the formalism developed in this section for connecting a global hybrid functional to its parent GGA.

\section{Specific cases} 
\label{sec_spec}

\subsection{Quantifying errors with DC-DFT}
\label{sec_exact}
There has been recent great interest in quantifying the errors in densities in DFT~\cite{MedBusSunPerLys-SCI-17,Kep-SCI-17,Ham-SCI-17,Kor-ANG-17}. However, in Ref~\cite{SimSuhBur-JPCL-18}, it was shown that the theory of DC-DFT provides a natural and unambiguous measure of density error.  With that measure, it was not possible to distinguish the densities of empirical and non-empirical functionals based on 
their self-consistent densities alone.

To understand the background to this, we first must distinguish ground-state KS
DFT from other areas of electronic structure.  The primary purpose of such calculations
is to produce the ground-state energy as a function of nuclear coordinates.  Indeed, 
in principle, one can deduce the density (and hence any integral over it) from a sequence
of such calculations, via the functional derivative with respect to the potential.  
Of course, such calculations produce KS  potentials, orbitals, and eigenvalues,
as well as densities and ground-state energies, and all such quantities can be compared
(for systems for which the calculation is feasible) to their exact counterparts extracted
from a more accurate quantum solver~\cite{UmrGon-PRA-94,ZhaMorParr-PRA-94,RyaKohSta-PRL-15}.   These are of great interest as inputs to response
calculations, such as in linear-response TDDFT or GW methods, and such procedures might
be extremely sensitive to such inputs.  But in ground-state DFT, the main prediction is
the energy of the many-body system, for which the KS scheme is simply a brilliant construct
that balances efficiency and accuracy.  

Intuitively, one feels that a 'better' XC potential must yield a 'better' density, and in turn,
a 'better' density must yield a better energy.   After all, the Hohenberg-Kohn (HK) theorem tells us that we reach
the ground-state energy only with the exact density and exact KS potential.  But such formal
statements give no measure of the quality of a density or a potential.  Even a well-defined 
mathematical norm between two densities that vanishes as the exact density is approached does
not really provide what we wish for, as an infinity of arbitrarily different norms can be
constructed.  All can tell us when we have found the exact density, but give differing results
for how far away we are from it.  A deep part of the problem is that both potentials and densities
are functions of $\br$, and so are not characterized by a single number.

As mentioned in Ref~\cite{SimSuhBur-JPCL-18}, the basic theorems of DFT give
us an ideal solution to this dilemma, via density-corrected DFT. 
To write this measure in the language of the density functional analysis, we consider now the specific case
where $E_{\xc}^{(0)}[n]=E_{\xc}[n]$ is the exact XC functional and where $ E_{\xc}^{(1)}[n]=\tilde E_{\xc}[n]$ is
an approximate functional. Note also that this is the basis of all DC-DFT applications. For a given $v(\br)$,
the difference between the two corresponding ground state energies becomes the error in the
approximate ground-state energy:
\beq
\Delta \tilde{E}_v= \tilde{E}_v-E_v.
\eeq
For this special case, Eq.~\ref{eq:dif2} becomes:
\begin{align} \label{eq:dif2exact}
\Delta \tilde E_v  =  \underbrace{
 \tE_v [\tilde{n}_v]- E_v [\tilde{n}_v]}_
{\Delta \tilde E_{\xc}[\tilde{n}_v]} + \underbrace{E_v [\tilde{n}_v]  - E_v [n_v]}_
{\Delta E^{ideal}[\Delta \tilde{n}_v]=D_v[\Delta \tilde{n}_v]}.
\end{align}
where $\Delta \tilde n_v = \tilde n_v - n_v$.
For any system, $\Delta E^{ideal}[\Delta \tilde{n}_v]$ is a positive energy for any $\tilde\n_v(\br)$ and
vanishes only for the exact density.  This choice is ideal because (a) there are no human choices within the measure to argue over, and (b) the measurement is
in terms of the energetic consequences.  Thus it even 
provides a scale for density differences.  For example, if this measure yields results in the microHartree
range, why would one even care about errors in the density? Given that it is very difficult in practice to evaluate the exact functional on an approximate density (but see e.g., Refs.~\cite{WSBW-PRL-13} and~\cite{locpaper}), DC-DFT procedures use the following equation instead:
\begin{align} \label{eq:dif1exact}
\Delta \tilde E_v =  \underbrace{
\tE_v [\tilde{n}_v]- \tE_v [n_v]
}_
{\Delta E_{\D}=-\tilde{D}_v [ -\Delta \tilde n_v]} 
+ \underbrace{ \Delta \tilde E_{\xc}[n_v]}_
{\Delta E_{\F}}.
\end{align}
\begin{figure}[htb]
\includegraphics[width=1.0\columnwidth]{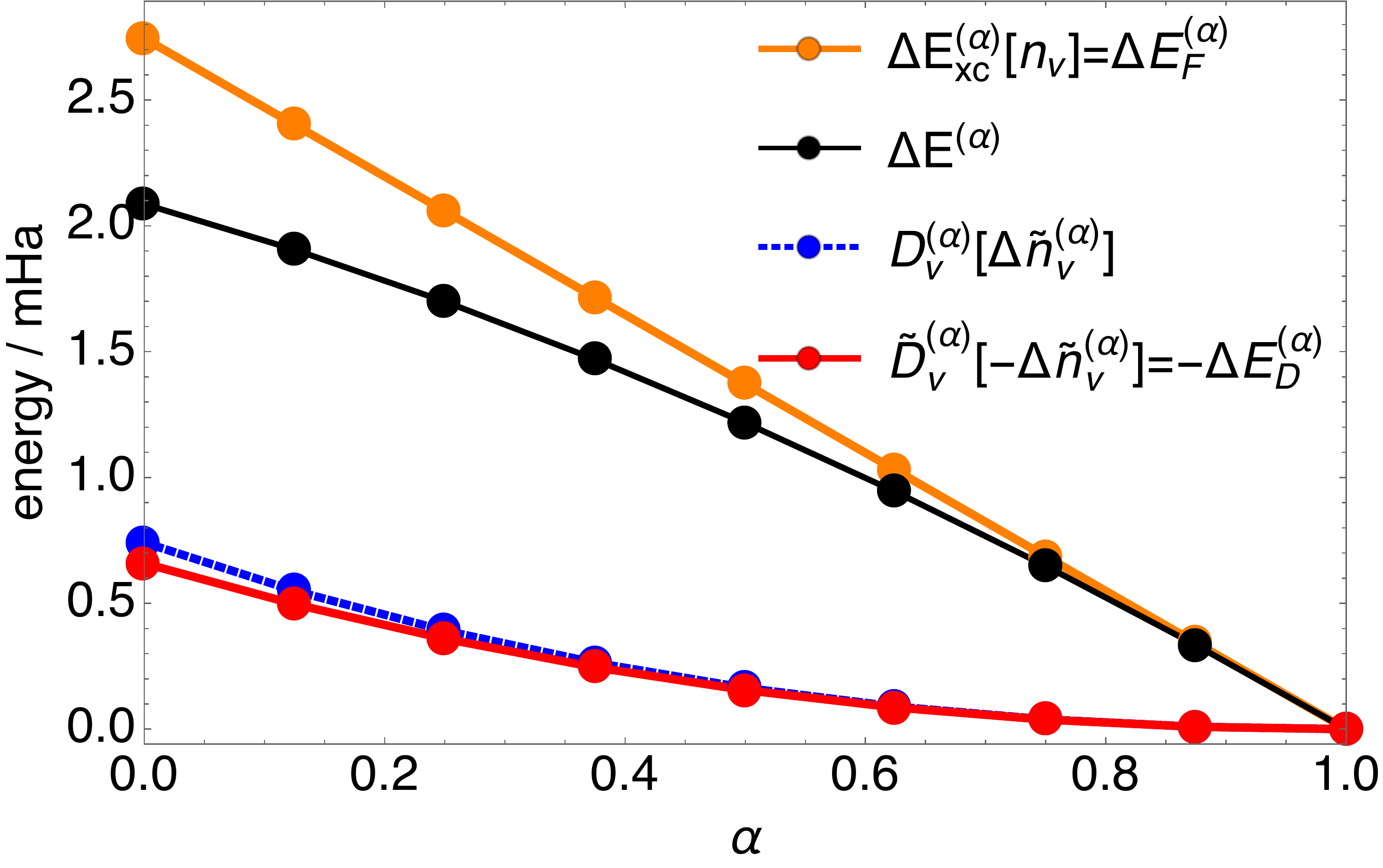}
\caption{Various errors of Eqs.~\ref{eq:dif1exactalpha} and~\ref{eq:dif2exactalpha} for the $\alpha$-BLYP calculations of the hydrogen atom as a function of amount of exact exchange mixing}
\label{fig:blyp_h}
\end{figure}
Equation~\ref{eq:dif1exact} allows us to decompose $\Delta \tilde E_v$, the total error made by $\tilde E_{\xc}[n]$ and $\tilde n_v$, into the
functional error $\Delta E_{\F}=\Delta \tilde E_{\xc}[n_v]$ and the density-driven error $\Delta E_{\D}=-\tilde{D}_v [ -\Delta \tilde n_v]$,
which is much more practical than the ideal, as it needs only be evaluated on the approximate functional.
We can in fact expect $\Delta E_{\D}$ to be a practical proxy for the intractable $E^{ideal}[\Delta \tilde{n}_v]$ measure. When the approximate density is sufficiently close to its exact counterpart (more precisely, when the two inequalities hold: $D_v[\Delta \tilde n_v] \leq \Delta_c$ and $\tilde{D}_v [ -\Delta \tilde n_v] \leq \Delta_c$), we can write $\Delta E^{ideal}[\Delta \tilde{n}_v]$ as:
\beq \label{eq:cataylorexact}
\Delta E^{ideal}[\Delta \tilde{n}_v] \approx    \half   K_v [\Delta \tilde n_v] ,
\eeq
and $\tilde{D}_v[-\Delta n_v]$ as: 
\beq \label{eq:cataylorapproximate}
\tilde{D}_v[-\Delta \tilde n_v]\approx   \half   \tilde K_v  [     \Delta \tilde n_v         ] 
\eeq
From Eqs.~\ref{eq:cataylorexact} and~\ref{eq:cataylorapproximate}, we can see that if the approximate functional has accurate curvature, $\tilde{D}_v[-\Delta \tilde n_v]= - \Delta E_{\D}$ measure  is very similar to $\Delta E^{ideal}[\Delta \tilde{n}_v] $. 

\subsection{Illustration}
Figure~\ref{fig:blyp_h} illustrates many aspects of the analysis described so far.
Here we consider the hydrogen atom and the BLYP GGA~\cite{Bec-PRA-88,LeeYanPar-PRB-88}. We choose this example carefully,
because (a) we have easy access to the exact density, since this is a one-electron case, 
and (b) our functional correctly has no correlation energy (as LYP
correlation vanishes for all fully spin-polarized systems).  Thus,
when we interpolate between BLYP and HF, we create a global hybrid
with a fraction $\alpha$ of exact exchange (EXX):
\beq \label{eq:exc_hyb}
\tilde E_{\xc}^{(\alpha)}[n]=\tilde E_{\xc}^{\GGA}[n]+ \alpha \left(E_{\x}[n]- \tilde  E_{\x}^{\GGA}[n]\right).
\eeq
For this specific cases $\tilde E_{\xc}^{(\alpha)}[n]$ reduces to:
\beq
\tilde E_{\xc}^{(\alpha)}[n]= \tilde E_{\x}^{\rm B88}[n]+ \alpha \left(E_{\x}[n]- \tilde  E_{\x}^{\rm B88}[n]\right),
\eeq
where $E_{\x}^{\rm B88}[n]$ stands for the exchange functional of Becke~\cite{Bec-PRA-88}.
For the $\Delta \tilde E_v^{(\alpha)}=\tilde E_v^{(\alpha)}-E_v$ energy difference (i.e. the error of the hybrids functional), we can rewrite Eq.~\ref{eq:dif1exact} as:
\begin{align} \label{eq:dif1exactalpha}
\Delta \tilde E_v^{(\alpha)}=\underbrace{-\tilde{D}_v^{(\alpha)} [ -\Delta \tilde n_v^{(\alpha)}]}_{\Delta E_{\D}^{(\alpha)}}
&+ \underbrace{ \Delta \tilde E_{\xc}^{(\alpha)}[n_v]}_
{\Delta E_{\F}^{(\alpha)}},
\end{align}
where $\Delta \tilde n_v^{(\alpha)}= \tilde n_v^{(\alpha)}-n_v$. 
In Eq.~\ref{eq:dif1exactalpha}, we can recognize $\alpha$-dependent density-driven and functional errors. In the same manner, we can re-write Eq.~\ref{eq:dif2exact}:
\begin{align} \label{eq:dif2exactalpha}
\Delta \tilde E_v^{(\alpha)}  = 
\Delta \tilde E_{\xc}^{(\alpha)}[\tilde{n}_v^{(\alpha)}]
+D_v[\Delta \tilde{n}_v^{(\alpha)}].
\end{align}
In this case, all error contributions (Eqs.~\ref{eq:dif1exactalpha} and~\ref{eq:dif2exactalpha}), shown in Figure~\ref{fig:blyp_h}, vanish at $\alpha=1$. 
On the extreme left ($\alpha=0$), we see the functional error exceeds the
self-consistent error, and the density-driven error is negative, as it
should be.  The functional error is exactly linear, going to
zero as $\alpha \to 1$.  Notice that the density-driven error must always
behave parabolically around $\alpha=1$.  

We also compare the two choices of reference for $D_v$ (blue and red), finding
that they are almost identical.  This is telling us that the BLYP density is
sufficiently close to the exact density that the expansion to second-order is
fine.  Moreover, note that as $\alpha \to 1$, the blue and red merge datapoints, and both
are on top of a perfect parabola whose curvature is given by $K_v [\Delta n]$ (again as  $\alpha \to 1$).
Since the
self-consistent error is the sum of the functional and density-driven errors, we
can deduce its curve just from the values at $\alpha=0$.   Thus the black line
is always a parabola if the densities are sufficiently similar, as is the case here.
Note that we can see that the energetic difference between the $D_v$ values (blue and red plots) is
rather low, showing that they are indeed sufficiently close that there are no significant
energetic consequences to approximating all such curves as parabolas.

Finally, we note that, relative to DC-DFT, we have chosen the two functionals in the
reverse order:  0 denotes the approximate functional, 1 the exact answer.  This is to 
make these results more readily comparable to other results for hybrids.   Simply
replace $\alpha$ by $1-\alpha$ to make Fig.~\ref{fig:blyp_h} in the form for DC-DFT.

\subsection{Self-interaction error and one-electron systems}
\label{sec_one}

For one-electron systems:
\begin{align}
 \label{eq:F1}
F[n]=T_{\s}[n],\,  E_{\x}[n]=-U_{\sss H}[n],\, E_{\c}=0~~~~~(N=1).
\end{align}
Standard DFT approximations typically do not satisfy all the conditions of Eq~\ref{eq:F1}, and for this reason, they suffer from 
one-electron self-interaction error (SIE)~\cite{CohMorYan-CR-12,PerZun-PRB-81}. On the other hand, the HF method is exact for one-electron systems, and thus
we can use the HF method to calculate the functional- and density-driven term of SIE. 
\begin{figure}[htb]
\includegraphics[width=1.0\columnwidth]{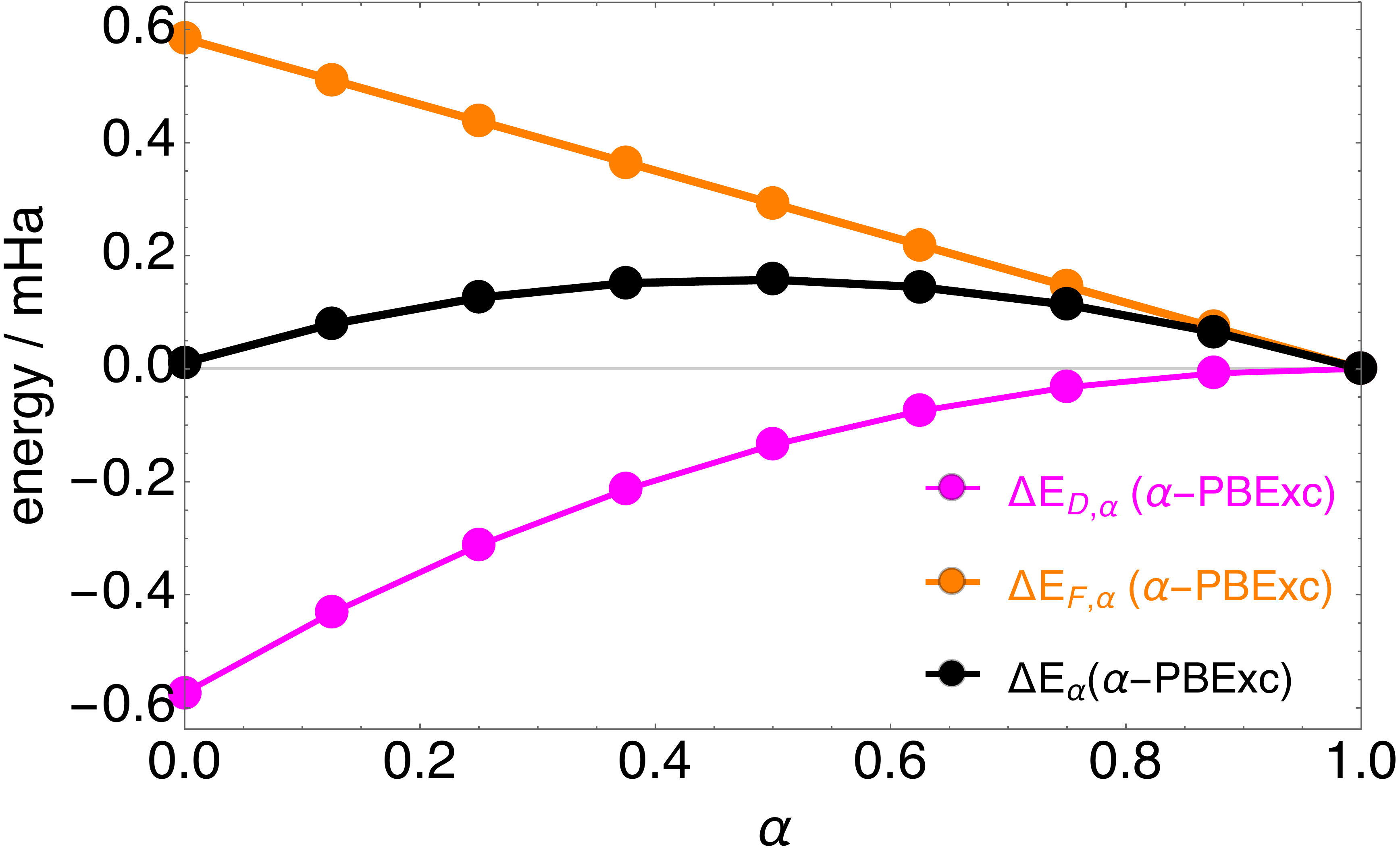}
\caption{Density-driven and functional errors for the $\alpha$-PBExc (Eq.~\ref{eq:exc_pbe}) calculations of the hydrogen atom as a function of amount of exact exchange mixing}
\label{fig:pbe_h}
\end{figure}
This has been already done in Figure~\ref{fig:blyp_h} for the BLYP hybrids, and 
in Figure~\ref{fig:pbe_h} we apply the same analysis to hybrids from the PBE functional~\cite{PerBurErn-PRL-96}. 
The PBE correlation energy, unlike that of LYP, does not vanish for one-electron systems.  
For this reason, in the case of the PBE functional we modify Eq.~\ref{eq:exc_hyb}:
\begin{equation}
\label{eq:exc_pbe}
\tilde E_{\xc}^{(\alpha)}[n]=\tilde E_{\xc}^{\PBE}[n]+
\alpha~(E_{\x}[n]-\tilde E_{\xc}^{\PBE}[n]),
\end{equation} 
In this way, we ensure that the error of the PBE hybrid of Eq.~\ref{eq:exc_pbe} (hereinafter $\alpha$-PBExc) vanishes at $\alpha=1$ 
Note that this does {\em not} include  PBE0~\cite{PerErnBur-JCP-96}, as this PBExc has only 0.75 of PBE correlation at $\alpha=1/4$.
In Figure~\ref{fig:pbe_h}, we show the density-driven  and functional error of $\alpha$-PBExc for the hydrogen atom. 
First, note that the scale of the errors is minuscule.
We can also see that both $|\Delta E_{\F}^{(\alpha)}|$ and $|\Delta E_{\D}^{(\alpha)}|$ errors of
$\alpha$-PBExc decrease with
$\alpha$. Nonetheless, its total $\Delta E^{(\alpha)}$ error peaks at $\alpha \approx 0.5 $ and nearly vanishes for
the PBE functional (the $\alpha = 0 $ case). The fact that the PBE gives almost exact energy for the hydrogen atom relies on 
a cancellation between the functional and density-driven errors (as well as a cancellation between exchange and correlation
errors).  

\begin{figure}[htb]
\includegraphics[width=1.0\columnwidth]{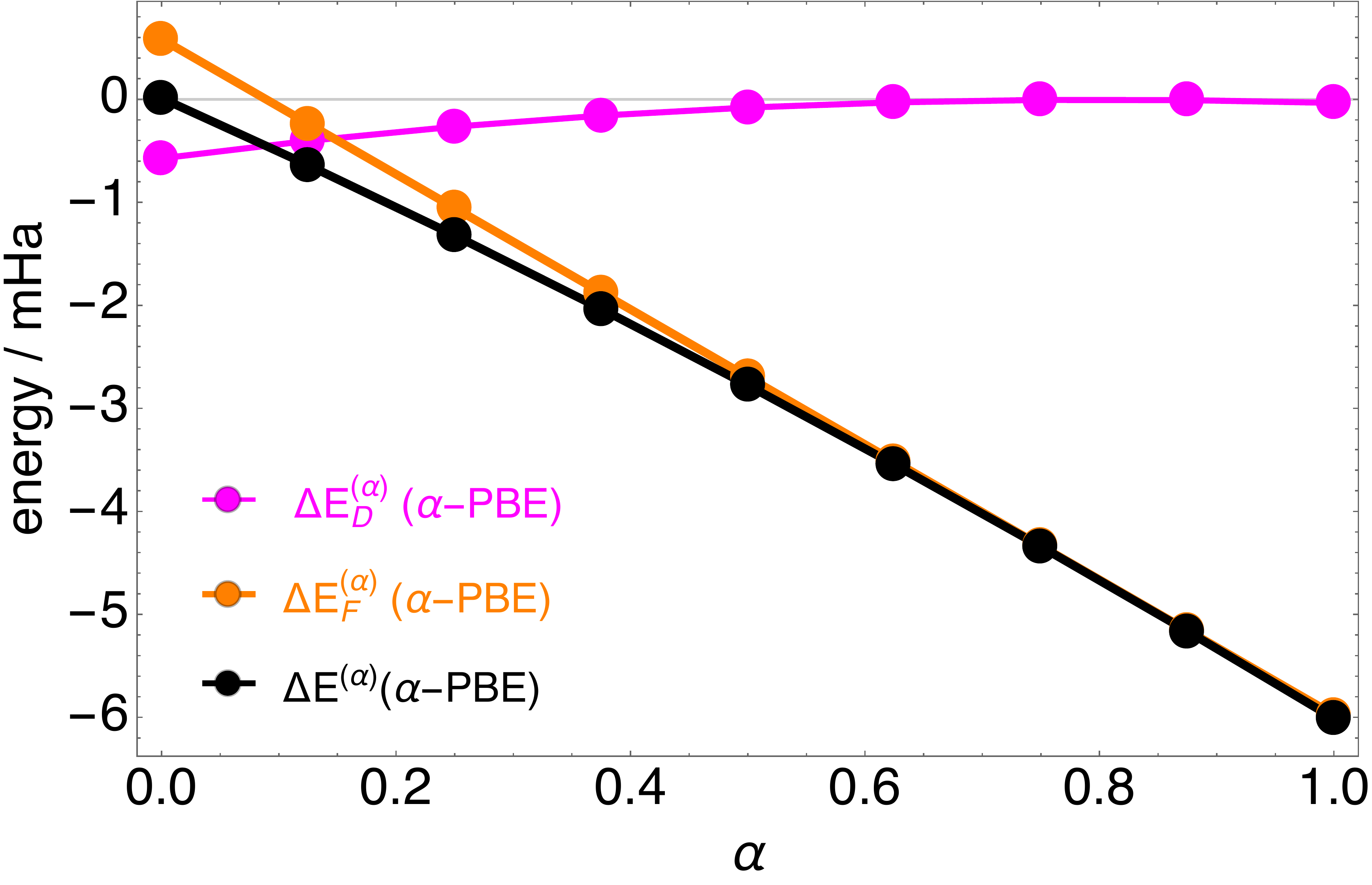}
\caption{Density-driven and functional errors for the $\alpha$-PBE (Eq.\ref{eq:exc_hyb}) calculations of the hydrogen atom as a function of amount of exact exchange mixing}
\label{fig:apbe_h}
\end{figure}
The same plot for the hydrogen atom obtained with the regular $\alpha$-PBE hybrid (Eq.~\ref{eq:exc_hyb}) is shown
in Figure~\ref{fig:apbe_h}. Note that now the $\alpha=1/4$  point represents the PBE0 functional. We can see from Figure~\ref{fig:apbe_h} that as $\alpha$ approaches $1$, the functional error strongly dominates its functional-driven counterpart. Note here the much larger scale:  The self-interaction error in the PBE correlation functional error yields much larger total energy
errors than in the previous figure, illustrating the increased error when semilocal correlation functionals are combined with exact
exchange. We can also note that $E_{\D}$ gets very close to $0$ as $\alpha$ approaches $1$, although the PBE correlation potential does not vanish for $N=1$ systems.


\subsection{The Hartree approximation} 

\begin{figure}[htb]
\includegraphics[width=1.0\columnwidth]{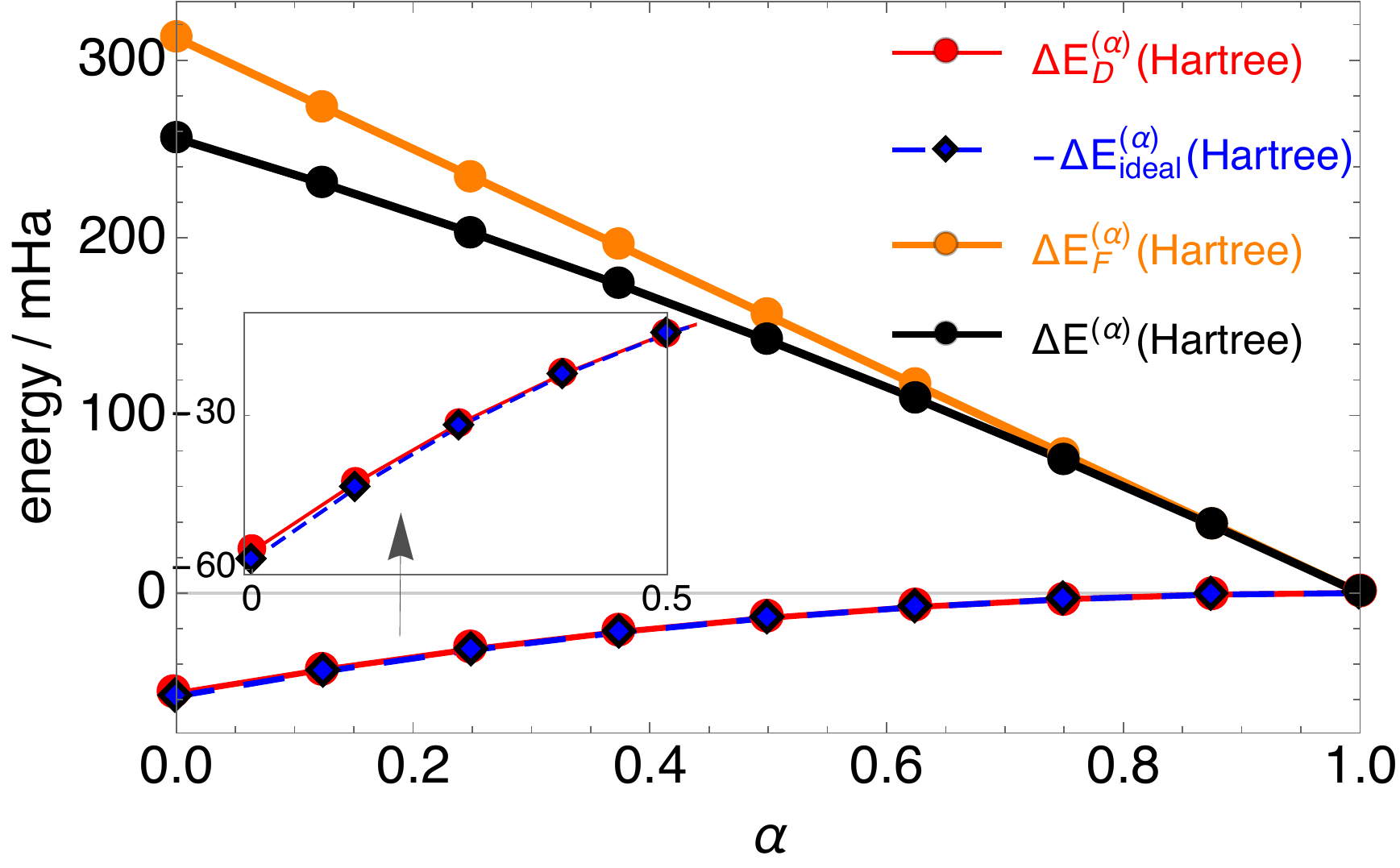}
\caption{Density-driven, the ideal, and functional errors for the hydrogen atom calculation with the $\tilde E_{\xc}^{(\alpha)}[n]=\alpha \tilde E_{\x}^{(\alpha)}[n]$ functional. 
}
\label{fig:hartree}
\end{figure}

Another special case is the Hartree approximation, i.e., solution of the KS equations
with XC set to zero.   Here, we compare any non-zero $\tilde E\xc$ with pure Hartree.
We choose Hartree to be $\alpha=0$, so that $\alpha$ then represents the fraction of 
$\tilde E\xc$ included in the calculation.  While Hartree calculations are certainly
too inaccurate for chemical purposes~\cite{PerSch-AIP-01,Per-MRS-13}, one would expect them to have the
greatest delocalization error of any approximate functional, since not even LDA exchange
is opposing the Hartree energy.  They might thus prove useful in creating a non-empirical
measure of delocalization to be used in DC-DFT.

If the reference is the exact XC energy, then the functional difference is just the
XC energy itself, while the density-driven error is just: $\Delta E_{\D}=-\half f_{\SH}[n_v^{(0)}, \Delta\n]$.
In this case, one would expect $E_{\D}$ to be different from the ideal, which includes the XC
contributions. However, as we will show below the two quantities are nearly the same for the hydrogen atom.

To give an illustration, we consider the following  functional: $\tilde E_{\xc}^{(\alpha)}[n]=\alpha E_{\x}[n]$, which for one-electron systems connects the Hartree approximation ($\alpha=0$) to  the exact functional ($\alpha=1$). In Figure~\ref{fig:hartree}, we show the errors of this functional as a function of $\alpha$ for the hydrogen atom. As expected, the scale of errors is much larger than those shown in Figures~\ref{fig:blyp_h}-\ref{fig:apbe_h}. 
We find it interesting that the $-\Delta E^{ideal}$ datapoints in  Figure~\ref{fig:hartree} are hardly distinguishable from their $\Delta E_{\D}$ counterparts. Thus, at $\lambda=0$ we have: 
\beq \label{eq:Hsh}
\underbrace{f_{\sss S}[n_v,\Delta\n]}_{-2\Delta E^{ideal}}    \sim 
\underbrace{ f_{\SH}[n_v^{(0)}, \Delta\n]}_{-2\Delta E_{\D}}~~~~
\rm{(H~atom)},
\eeq
where $f_{\sss S}(\br,\br')$ is the kinetic component of $ f_{\SH}(\br,\br')$ (see Eq.~\ref{eq:s}). 

\subsection{Pure density functionals}

\begin{figure}[htb]
\includegraphics[width=1.0\columnwidth]{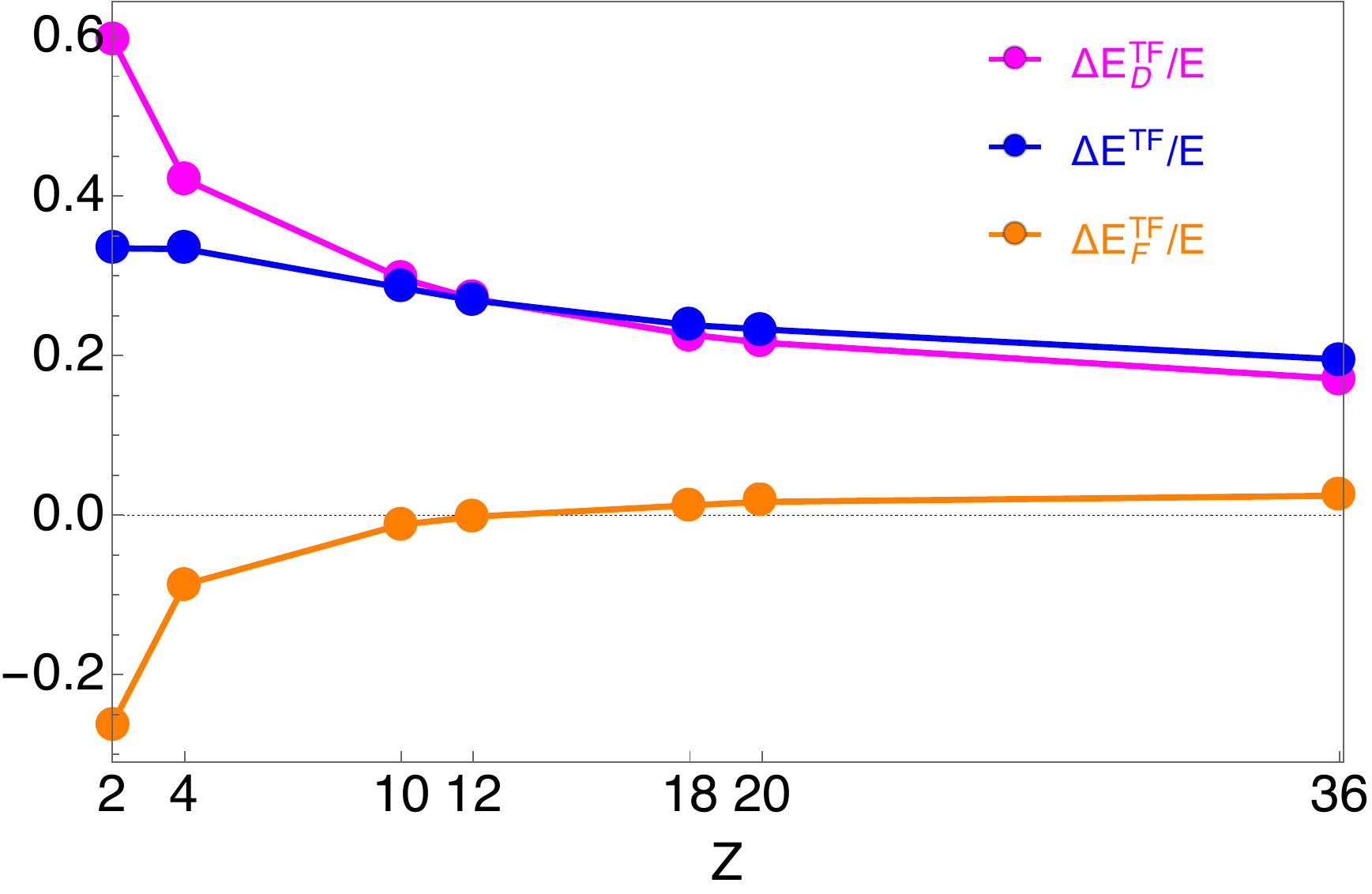}
\caption{Plots showing quantities that involve the density-driven and functional errors of the Thomas-Fermi method (Eq.~\ref{eq:tfdf}) with $Z$ for a range of small atoms. 
}  
\label{fig:tf}
\end{figure}

So far, we have considered only approximations to XC within the KS scheme, as this is the
most common DFT calculation today by far.   However, there is much interest in developing
orbital-free functionals, especially in contexts when the KS scheme becomes too cumbersome.

Since the entire functional $F[\n]$ is approximated in such a scheme, the density is often
much poorer than in a KS calculation.  In fact, estimates suggest that simple orbital-free
approximations, such as those used in Thomas-Fermi theory~\cite{Tho-MP-27,Fer-1927,LieSim-AIM-77}, produce sufficiently poor
densities that their errors are {\em dominated} by errors in the density, i.e., the density-driven
error is much larger than the functional error in most calculations.   This is seen in total
energy calculations of atoms and of one-dimensional fermions in a flat box~\cite{WasNafJiaKimSimBur-ARPC-17}.  The simplest
DC-DFT in orbital-free DFT is to apply the approximation on the exact density to eliminate
the density-driven error.

To exemplify the error analysis of orbital-free functionals, we consider here the TF energy functional, whose universal part reads as:
\begin{align}
F^{\TF}[n]=
\underbrace{T_{\s}^{\TF}[n]}_{A^{\TF}  \intr n(\br)^{5/3} } + U[n],
\end{align}
with $A^{\TF}=\frac{3}{10}\left( 3 \pi^2\right)^{\frac{2}{3}}$.  The total TF error can be, analogously to eq~\ref{eq:dif1exact}, partitioned as:
\begin{align} \label{eq:tfdf}
\Delta  E_v^{\TF} =  \underbrace{
E_v^{\TF} [n_v^{\TF}]- E_v^{\TF} [n_v]
}_{\Delta E_{\D}^{\rm TF}}
+
\underbrace{\Delta F^{\rm TF} [n_v] }_
{\Delta E_{\F}^{\rm TF}},
\end{align}
where $\Delta F^{\rm TF} [n]= \Delta E_{v}^{\rm TF}[n]= \Delta T_{\s}^{\TF}[n]-E_{\xc}[n]$. 
Here we calculate $\Delta E_{\D}^{\rm TF}$ and $\Delta E_{\F}^{\rm TF}$ for alkaline earth metals and noble gases up to krypton (Z=36). They are computed by utilizing that for neutral atoms: $E_{Z}^{\rm TF} \sim -0.7687  \, Z^{7/3}$~\cite{ParYan-BOOK-89}. Highly accurate energies and densities, i.e., $E_v$ and $n_v$ entering Eq.~\ref{eq:tfdf} have been obtained from the PySCF software~\cite{PYSCF} at the CCSD level within the aug-cc-pV$m$Z basis set~\cite{Dun-JCP-89}, with the largest $m$ available for each of the atoms.
From the plots shown in Figure~\ref{fig:tf}, we can see that the density-driven component strongly dominates the total $\Delta E^{\rm TF}$ error. For example, in the case of the neon atom ($Z=10$) most of the TF error is practically density-driven, with $ E_{\D}^{\rm TF} / E  \sim 28.4 \%$  and $ E_{\F}^{\rm TF} / E $ being only $-1.3 \%$! We remember that for neutral atoms, as $Z \to \infty$, the TF theory becomes relatively exact, in the sense that it satisfies:~\cite{LieSim-PRL-73,BurCanGouPit-JCP-16,CanChenKruBur-JCP-18}
\begin{align} \label{eq:Ztf}
\lim_{Z \to \infty} \frac{\Delta E_Z^{\rm TF}}{E_Z}   \to 0.
\end{align}
Thus, as $Z \to \infty$,  our blue curve in Figure~\ref{fig:tf} should vanish. Nevertheless, in Figure~\ref{fig:tf} we are still far from this limit, as at our largest $Z$ value ($Z=36$), $\Delta E_Z^{\rm TF}/ E_Z$ is around $1/5$.

This suggests several important points regarding these functionals. First, they must always
be tested self-consistently, as tests of new orbital-free approximations on KS densities does not tell us much about their overall accuracy, given that the density-driven errors can be very large. At the same time, comparison of the functional on the KS density then provides
enough information to separate functional- from density-driven errors,  and we expect that even the KS densities obtained from the (semi)local XC approximations are sufficiently accurate for this purpose. Second, reports of
failures of TF theory and its extensions should be revisited to determine if these are density-driven or functional-driven.
If the former, one should focus on improving the densities rather than the total energies alone.
Third, this supports efforts~\cite{Fin-JCP-16} to approximate the Pauli potential~\cite{Mar-PLA-86,LevOuy-PRA-88} directly as a density functional, without requiring that the KS potential be a functional derivative.

In the context of the present paper, it should prove useful when comparing two orbital-free
approximations to decompose their differences into functional- and density-driven contributions.
If two different approximations differ in both contributions, this would suggest that good
aspects of both might be combined to separately minimize each error.

\subsection{Strong correlation}
\label{strong_corr}

In our last example, we show that density-driven errors can become
large when systems are strongly correlated, but need not be.  Generating
a simple example is not so easy, as one needs essentially exact densities
upon which to make evaluations and comparisons.   
Fortunately, the two-site two-fermion
Hubbard model is an example where all quantities
can be determined analytically.  Many relevant KS-DFT quantities have
been calculated exactly and summarized in two recent reviews, one on the
ground-state\cite{CarFerSmiBur-JCP-15} and one on linear-response
TDDFT.\cite{CarFerMaiBur-EPJB-18} 

For any two-electron system (in the absence of magnetic fields), the
restricted HF functional is:
\ben
F^{RHF}[n]=T\s[\n] + U_{\sss H}[n]/2,
\een
as half the Hartree is canceled by exchange, and correlation is ignored.
In fact, the traditional definition of correlation energy in quantum 
chemistry, is
\ben \label{eq:EQC}
E\c^{QC}=E-E^{RHF}[\n^{RHF}].
\een
Thus the energy-driven error of RHF is:
\ben \label{eq:EF}
\Delta E_F = - E\c,
\een
while the density-driven error is
\ben \label{eq:ED}
\Delta E_D = E\c - E\c^{QC}.
\een
For our 2-site model, the functional reduces to a simple function.
The onsite occupations are 
$n_1$ and $n_2$, i.e., the density is just two non-negative real numbers.  Moreover,
since they always sum to 2, the density is fully represented by their
difference.  Likewise, we can choose the average potential to be zero and
represent the inhomogeneity in the potential by a single number, $\Delta v$,
the on-site potential difference.  If we choose the hopping parameter $t=1/2$,
the only other parameter is $U$, the energy cost of double occupation of a site.

\begin{figure}[htb]
\includegraphics[width=1.0\columnwidth]{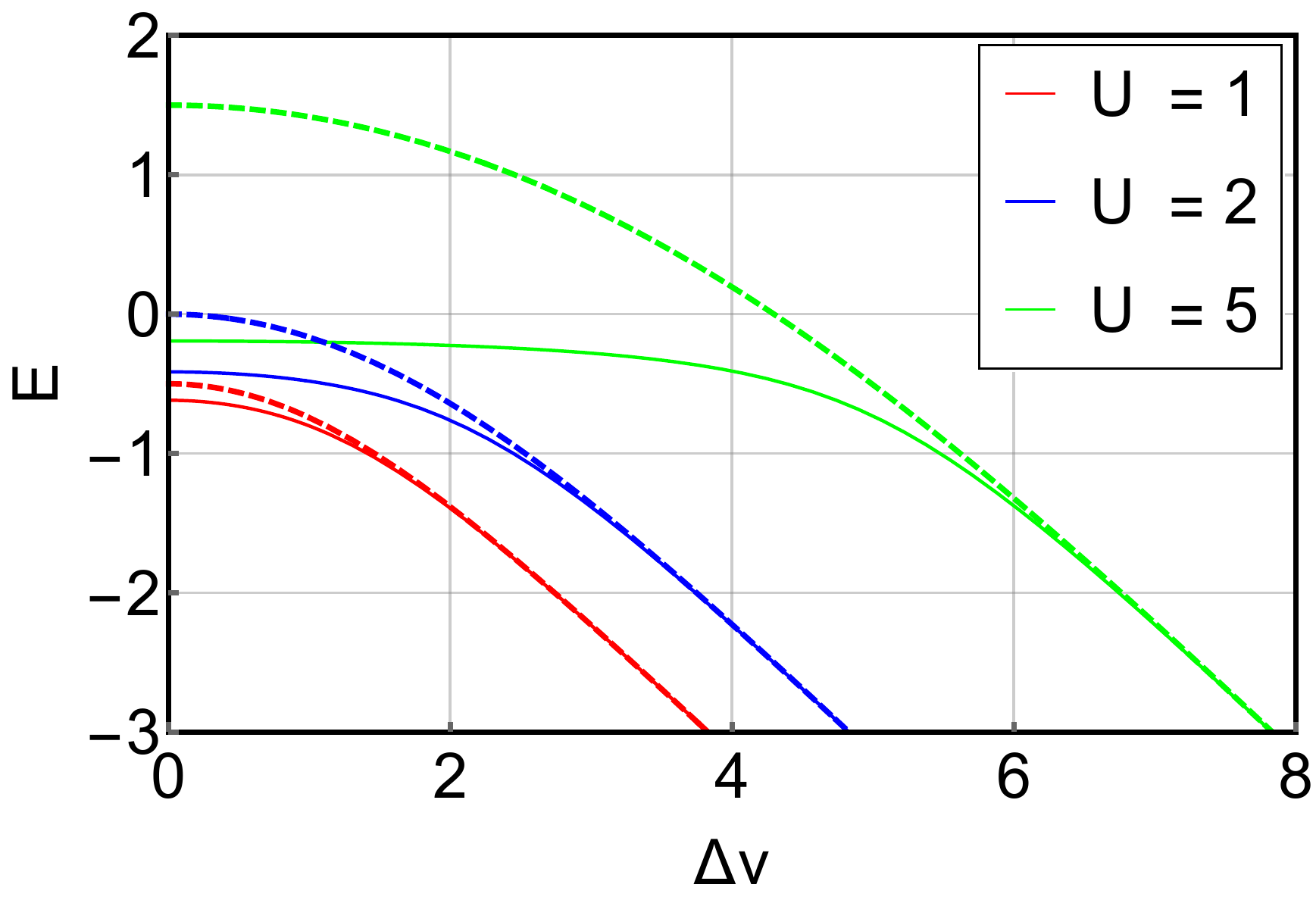}
\caption{Restricted Hartree-Fock Hubbard dimer ground-state energy (dashed line)
and exact Hubbard dimer ground-state (solid line) as
functions of $\Delta v$ for varying values of $U$ (see Ref.\cite{CarFerSmiBur-JCP-15}}
\label{fig:HDEnergy}
\end{figure}

The error in RHF and the exact ground state energy is explored in Figure~\ref{fig:HDEnergy} for varying levels of correlation and
inhomogeneity.  The absolute error increases with $U$, as expected.
The energy error in energy for each level of correlation is most
prominent in the symmetric dimer $(\Delta v = 0)$, and diminishes
and rapidly vanishes beyond $\Delta v$ larger than $U$,
where the energy becomes linearly correlated with the
on-site potential difference. Thus the system becomes weakly
correlated for
$\Delta v > U$.

\begin{figure}[htb]
\includegraphics[width=1.0\columnwidth]{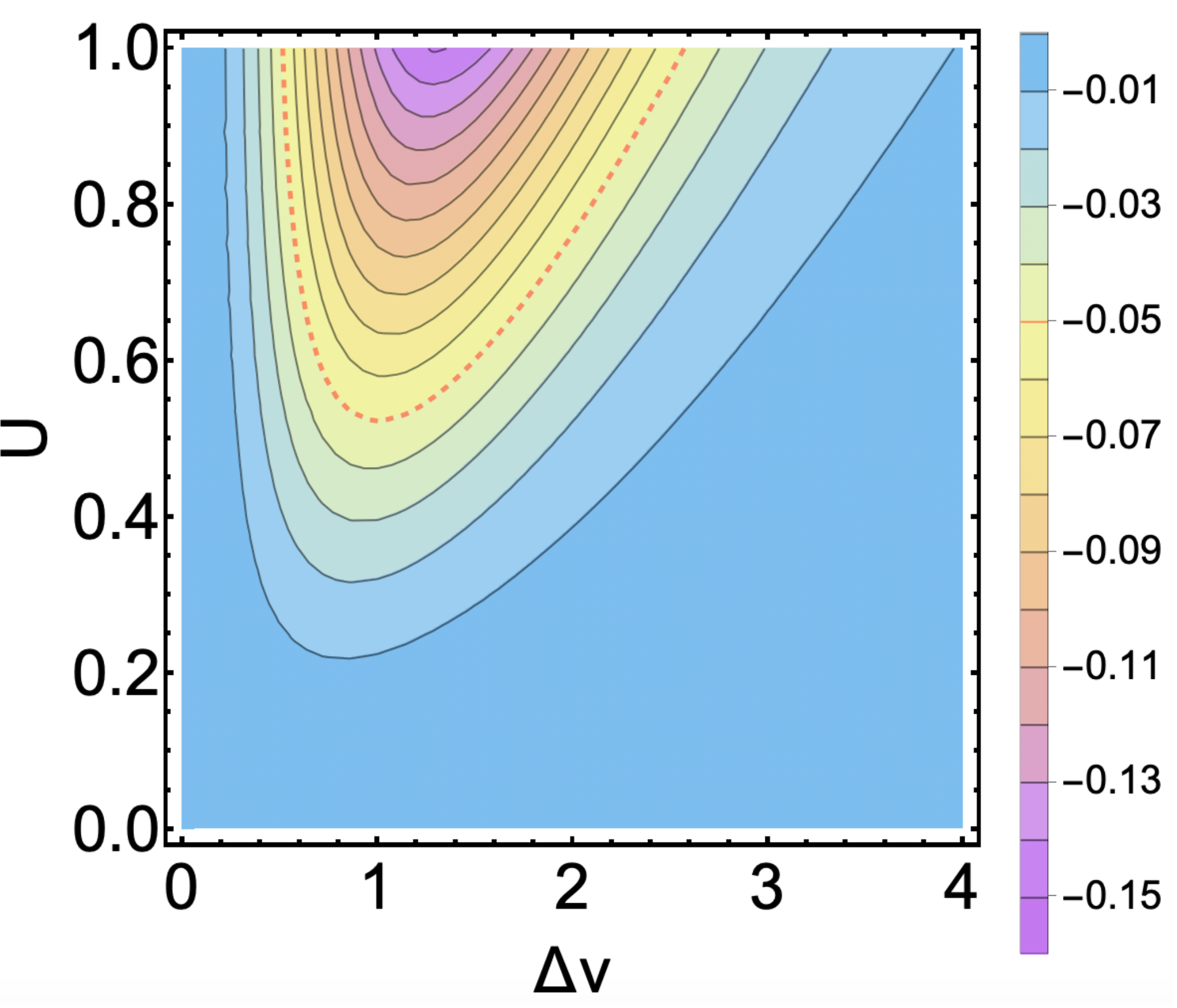} 
\caption{Fraction of error that is density-driven for moderate values of $U$,
with the 5\% contribution contour marked by a red, dashed line.}
\label{fig:contour1}
\end{figure}

The functional-driven and density-driven contributions to
this HF error were then isolated through the use of Eqs.~\ref{eq:EQC},~\ref{eq:EF}
and~\ref{eq:ED}. The fraction of the total error attributed to
the density-driven component $(\Delta E_D / E_C^{QC})$ is shown
in Figure~\ref{fig:contour1} for weakly correlated dimers with
values of $U$ up to 1. As $U$ decreases in size, both the 
magnitude of total error and its density-driven contribution decrease
substantially. For $U < 0.5$, there is no $\Delta v$ for which
there is a density-driven contribution greater than 5\% of the total error.
Of course, as $U\to 0$, this ratio must vanish, so this is not unexpected.
But we also see that the density-driven error vanishes at $\Delta v=0$
for any value of $U$, no matter how large, for symmetry reasons.  Thus
even a strongly correlated system might have no density-driven error.
Moreover, 
for $\Delta v > 1 + 2 U$, again the error is less than 5\%, due to correlation
being weakened by inhomogeneity.   So for any given $U$, there is a maximum
in the fraction of density-driven error as a function of $\Delta v$, and it 
is at non-zero $\Delta v$.

\begin{figure}[htb]
\includegraphics[width=1.0\columnwidth]{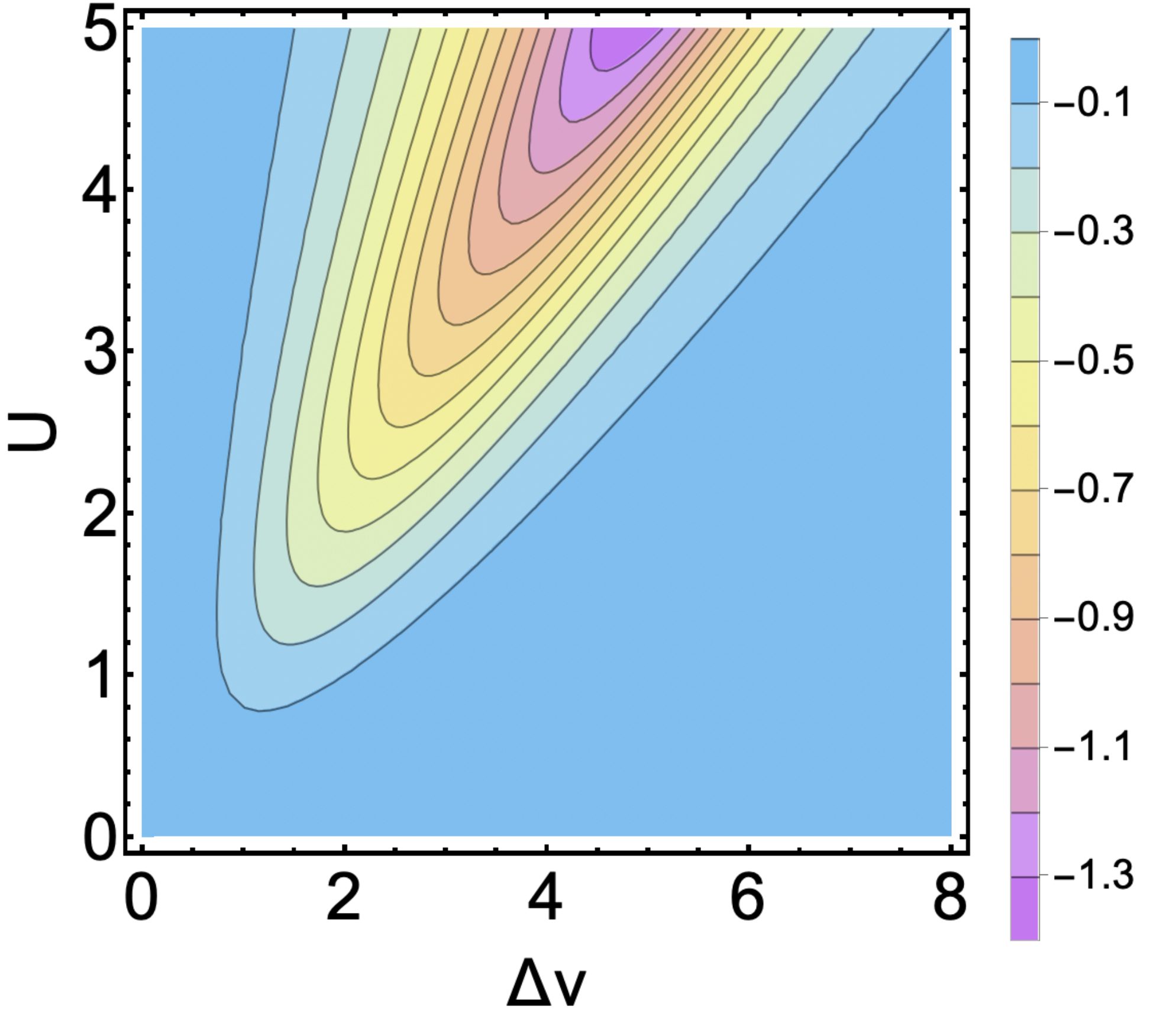}
\caption{Same as Fig. \ref{fig:contour1}, but zoomed out in $U$ and $\Delta
v$.  Note the change in contour/color scale.}
\label{fig:contour5}
\end{figure}

The density-driven error ratio for more strongly
correlated dimers is shown in Figure~\ref{fig:contour5}, and has
characteristics identical to the weak correlation plot, but
on a larger scale.  Clearly the maximum fractional density-driven error 
becomes much larger with $U$ and can even exceed -$1$.  We also see that
for $U > 1$, the region of small density-driven error around $\Delta v = 0$
can even increase with $U$.  For fixed $\Delta v$, the
fraction of density-driven error goes down with sufficiently large $U$!.
The relation between RHF density-driven error and strong correlation is
clearly not trivial.

To avoid confusion, we note that
this section has focused on the density-driven error 
in RHF.   In the more realistic calculations of weakly-correlated systems
in the rest of this work,
we often assume that error is much smaller than the density-driven error
of a semilocal DFT calculation, and hence HF-DFT yields more accurate
energies in such cases.  Because the Hubbard dimer is a site model, there 
is no genuine correspondence with semilocal DFT approximations to test on
here.

\section{Energy differences}
\label{sec_dif}

Key chemical concepts are determined by 
energy differences (e.g., atomization energies, ionization energies, barrier heights, reaction energies, etc.).
For this reason, we extend our analysis to energy differences. For simplicity, we first look at the energy difference of systems A and B, 
whose external potentials are $v_A$ and $v_B$, respectively. This energy difference obtained from a total energy functional that corresponds to
a given $E_{\xc}^{(j)}$ is given by: 
\beq \label{eq:EAB}
E_{AB}^{(j)}=E_A^{(j)} [n_A^{(j)}]- E_B^{(j)} [n_B^{(j)}].
\eeq
When two functionals are involved, we can also define the difference between $E_{AB}^{(1)}$ and $E_{AB}^{(0)}$:
\beq \label{eq:difAB}
\Delta E_{AB}=E_{AB}^{(1)} - E_{AB}^{(0)}.
\eeq
Plugging Eq.~\ref{eq:EAB} into Eq.~\ref{eq:difAB} gives:
\beq \label{eq:ab01}
\Delta E_{AB}=
E_A^{(1)} [n_A^{(1)}]
-E_A^{(0)} [n_A^{(0)}]
- \Big(E_B^{(1)} [n_B^{(1)}]
-E_B^{(0)} [n_B^{(0)}]
\Big).
\eeq
Plugging Eq.~\ref{eq:dif1} into Eq.~\ref{eq:ab01}, we can obtain the counterpart of Eq.~\ref{eq:dif1} for the energy differences between systems A and B:
\begin{align}
 \label{eq:dif1abR}
\Delta E_{AB}=&
\Delta E_{\xc}[n_A^{(0)}]-E_{\xc}[n_B^{(0)}] \non
&\underbrace{-D_A^{(1)}[-\Delta n_A]+ D_B^{(1)}[-\Delta n_B]}
_{-D^{(1)}_{AB}}.
\end{align}
Similarly, we can also plug Eq.~\ref{eq:dif2} into Eq.~\ref{eq:ab01}, to obtain the counterpart of Eq.~\ref{eq:dif2} for the energy differences between systems A and B:
\begin{align}\label{eq:dif2abR}
\Delta E_{AB}=&
\Delta E_{\xc}[n_A^{(0)}]
-\Delta E_{\xc}[n_B^{(0)}] \non
&+\underbrace{D_A^{(0)}[\Delta n_A]-D_B^{(0)}[\Delta n_B]}_{D^{(0)}_{AB}}.
\end{align}
In Eqs.~\ref{eq:dif1abR} and~\ref{eq:dif2abR}, we recognize $D^{(1)}_{AB}$ and $D^{(0)}_{AB}$ as the density-driven terms, pertinent to the energy differences between systems A and B. 
While $D_v^{(j)}$ of Eq.~\ref{eq:cdelta}, which corresponds to the total energies is always greater or equal to $0$, its counterpart that pertains to the energy differences (Eqs.~\ref{eq:dif1abR} and~\ref{eq:dif2abR}) does not have a definite sign. Furthermore, if we look at $D^{(0)}_{AB}=D_A^{(0)}[\Delta n_A]-D_B^{(0)}[\Delta n_B]$ (Eq.~\ref{eq:dif2abR}), where  $D_A^{(0)}[\Delta  n_A] \geq 0$ and $D_B^{(0)}[\Delta  n_B] \geq 0$ we can see that $D^{(0)}_{AB}$ can easily vanish when $D_A^{(0)}[n_A^{(1)}] \sim D_B^{(0)}[n_B^{(1)}]$. Therefore, $D^{(0)}_{AB}$ and $D^{(1)}_{AB}$ can vanish even when $n_A^{(1)}$ and $n_B^{(1)}$ are drastically different from $n_A^{(0)}$ and $n_B^{(0)}$, respectively. 

The shown example that involves the energy difference between systems A and B, can be easily generalized to any energy difference of interest. For instance, consider the following chemical reaction: 
\begin{align}
\sum_{l=1}^{L}{R}_l \to \sum_{m=1}^{M}{P}_m,
\end{align}
where $\{R_l\}$ is a set of reactants and $\{P_m\}$ is a set of products. Then the energy of this reaction obtained from the $E_v^{(j)}[n]$ functional is:
\beq \label{eq:ed}
E_{\rm ED}^{(j)} =\sum_{m=1}^{M}  E_{P,m}^{(j)} [n_{P,m}^{(j)}]
-                              \sum_{l=1}^{L}  E_{R,l}^{(j)} [n_{R,l}^{(j)}]. 
\eeq
The corresponding difference in $E_{\rm ED}^{(j)}$ between $j=0$  and $j=1$ functional is :
\beq  \label{eq:deled}
\Delta E_{\rm ED} = E_{\rm ED}^{(1)} - E_{\rm ED}^{(0)}.
\eeq
Then $D_{\rm ED}^{(0)}$ that corresponds to $\Delta E_{\rm ED}$ is given by:
\begin{align} \label{eq:c0}
D_{\rm ED}^{(0)}=\sum_{m=1}^{M}    D_{P,m}^{(0)}[\Delta n_{P,m}]
-\sum_{l=1}^{L}    D_{R,l}^{(0)}[\Delta n_{R,l}],
\end{align} 
and its $D_{\rm ED}^{(1)}$ counterpart is given by:
\begin{align} \label{eq:c1}
D_{\rm ED}^{(1)}=\sum_{m=1}^{M}    D_{P,m}^{(1)}[-\Delta n_{P,m}]
-\sum_{l=1}^{L}    D_{R,l}^{(1)}[-\Delta n_{R,l}].
\end{align}

Now that we established the density-driven terms  of any energy difference of interest, we ask the following key question: how can we quantify the abnormality character of a given property/system obtained from a pair of functionals? Along these lines, we first define:
\beq \label{eq:max}
\Delta_n=\max{ ( |D^{(0)}| ,|D^{(1)} | ) }.
\eeq
One would naturally think of the following indicator of abnormality:
\beq \label{eq:eta}
\eta=\frac{\Delta_n}{\left | \Delta E_{\rm ED} \right |}.
\eeq
Following this indicator, the abnormality character of a given property of interest increases with $\eta$. However, the $\eta$ indicator would be problematic when $\left | \Delta E_{\rm ED} \right |$ is small, and such properties/system will always be abnormal by default. To fix this problem, we introduce the {\em abnormality scale}. For a chosen property of interest (e.g., a barrier height involving organic molecules) we calculate the abnormality scale by using a dataset with similar systems/properties (e.g., other similar barrier heights involving organic molecules). We then set the abnormality as the root-mean-square of $K$ datapoints (i.e. $\Delta E_{\rm ED}$ energies) that form the dataset:
\beq 
\bar \Delta_{\rm ED} =\sqrt{\frac{1}{K}\sum_{k=1}^{K} \left({\Delta E_{{\rm ED},k}}\right)^2 },
\eeq
Therefore,  $\bar{\Delta_{\rm ED}}$ determines our abnormality scale, and it is specific to a given class of properties/system and to a given pair of energy functionals. Finally, we use $\Delta$ to define our abnormality indicator:
\beq \label{eq:ind}
\bar \eta= \frac{\Delta_n}{\bar \Delta_{\rm ED}}
\eeq
Also in the case of this indicator, the abnormality increases with $\bar  \eta$. To classify systems as normal or abnormal (again for a given pair of functionals) based on the $\bar \eta$ indicator, we introduce a cut-off, e.g., $\bar \eta_c \sim \frac{1}{3}$, and consider abnormal systems/properties those that have $\bar \eta$ greater than $\bar \eta_c$. The primary goal of the present work is to outline the theory behind the density functional analysis. Therefore, showing numerical examples where we will illustrate (ab)-normality of different properties for different pairs of DFT approximations, we leave for future work. 

Returning to DC-DFT, where functional 1 is some approximation and 0 is exact,
we do not of course have access to $D^{(0)}$ in order to calculate $\bar \eta$,
so we settle for $|D^{(1)}|$ alone, which is simply the usual density-driven error.
But as noted in Ref.~\cite{SimSuhBur-JPCL-18}, that error would usually require knowing the exact
density, which is usually either unavailable or unaffordable.  (In normal systems, the
HF density is NOT more accurate than the self-consistent density, and so cannot be used).
A generic, practical workaround for standard approximations for use in HF-DFT
is to define the density sensitivity
of a functional as~\cite{SimSuhBur-JPCL-18}:
\beq
\tilde E_{\sss DS} = \left | \tilde E [n_{\sss LDA}]-\tilde E [n_{\sss HF}] \right|,
\eeq
which should always estimate $|D^{(1)}|$ when $|D^{(1)}|$ is a significant fraction of $\Delta$.  Thus the PBE
sensitivity is plotted in Fig 4 of Ref.~\cite{KimSonSimBur-JPCL-19}, and averaged sensitivities were used in Table I of Ref.~\cite{SimSuhBur-JPCL-18}.

\section{Conclusions}

We have given a detailed account of the considerations that led to the
recent successes of density-corrected DFT.   We have also generalized
that theory to allow different approximate functionals to be compared
in the same way as DC-DFT allows one approximation to be compared with
exact results.  We have shown that typical density differences between
reasonably accurate functionals can be shown quantitatively to be close
enough to allow treatment via density functional analysis expansions 
truncated at second order. We consider different special cases of our analysis, 
and note that DC-DFT is just one of these. 
For a given pair of density functional approximations, we also construct measures for their {\em relative abnormality} (i.e. the situation when the density-driven terms strongly dominate the energy difference obtained with two different approximations). 

We have noted many pioneering efforts in the chemistry literature in which
density-corrected calculations were performed, usually based on intuition.
We also point out that, as long ago as 1996, Levy and G{\"o}rling advocated the
use of self-consistent exact-exchange calculations, with a correction defined
to produce the exact ground-state energy as a functional of the `wrong' 
EXX density~\cite{LevGor-PRA-95}.  For purposes of calculating energies, the approach of Levy and G{\"o}rling  is nearly equivalent to
HF-DFT, given that the EXX and HF densities are probably indistinguishable.  Of course, we only advocate this procedure for abnormal systems, as in normal systems we expect the density from the standard semilocal approximations to be more accurate than that of HF. 

Finally, we note that of course it is highly unsettling to run KS calculations
in this non-self-consistent fashion.  Many advantages that are often taken for
granted, such as the exactness of the Hellmann-Feynman theorem in the basis-set
limit, are no longer true, and many corrections need to be coded.  But, in fact,
all information about the density can be extracted from a sequence of total energy
calculations, since:
\beq \label{eq:nbr}
n(\br) = \frac{\delta E_v}{\delta v(\br)} .
\eeq
Treating HF as EXX, this leads to a predicted change in the density of
a HF-DFT calculation relative to a self-consistent DFT density:
\beq
\Delta n(\br) = \frac{\delta (E^{DFT}[\n\HF]-E^{DFT}[n^{DFT}])}{\delta v(\br)}.
\eeq
Thus, in principle, one could calculate the improvement to the density predicted by HF-DFT,
and any other properties depending only on the density. 

\vspace{5mm}
\begin{acknowledgement}

KB and JK acknowledge funding from NSF (CHE 1856165). 
SV acknowledges funding from the Rubicon project (019.181EN.026), which is financed by the Netherlands Organisation for Scientific Research (NWO). 
SS and ES were supported by the grant from the Korean Research Foundation (2017R1A2B2003552).

\end{acknowledgement}

\bibliographystyle{unsrt}
\bibliography{all}

\label{page:end}
\end{document}